\PassOptionsToPackage{svgnames,usenames,dvipsnames}{xcolor}
\documentclass[acmsmall,screen]{acmart}
\usepackage{graphicx}

\def\cpc#1{}

\def\itemname#1{\textbf{#1}}
\def\inlinecode#1{\lstinline[language=coq]{#1}}
\def\inlinejs#1{\lstinline[basicstyle=\small\ttfamily]{#1}}
\def\inlinespec#1{\lstinline[basicstyle=\small\ttfamily]{#1}}
\usepackage{tikz}
\usetikzlibrary{positioning}
\usepackage{forest}
\usepackage{subcaption}

\usepackage{listings}
\usepackage{lstcoq}

\usepackage[inline]{enumitem}
\usepackage{hyperref}
\usepackage{cleveref}
\usepackage{wrapfig}
\usepackage{proof}

\usepackage{pifont}
\newcommand{\cmark}{\ding{51}}
\renewcommand{\epsilon}{\varepsilon}

\Crefname{figure}{Figure}{Figures}
\crefname{figure}{Figure}{Figures}

\newlist{commalist}{description*}{4}
\setlist[commalist]{itemjoin={{,}},afterlabel=\unskip{{~}},before=\kern-0.3em,mode=unboxed}

\newcommand{\sstep}[1]{\texttt{#1}}

\newcommand{\ch}[1]{\texttt{#1}}
\newcommand{\str}[1]{``\texttt{#1}''}
\newcommand{\charclass}[1]{[#1]}
\newcommand{\ncharclass}[1]{\charclass{\text{\textasciicircum{}}#1}}
\newcommand{\dotc}[0]{\ensuremath{\cdot}}
\newcommand{\escaped}[1]{\text{\textbackslash{}}#1}
\newcommand{\disj}[0]{\mathbin{|}}
\renewcommand{\star}[0]{^{*}}
\newcommand{\plus}[0]{^{+}}
\newcommand{\lazystar}[0]{^{*?}}
\newcommand{\lazyplus}[0]{^{+?}}
\newcommand{\question}[0]{^{?}}

\newcommand{\rrep}[2]{^{\{#1,#2\}}}
\newcommand{\lstar}[0]{^{*?}}
\newcommand{\lplus}[0]{^{+?}}
\newcommand{\lquestion}[0]{^{??}}

\newcommand{\lrrep}[2]{^{\{#1,#2\}?}}
\newcommand{\pgroup}[1]{({?}{:}\ #1)}
\newcommand{\group}[2][]{(\ifthenelse{\equal{#1}{}}{}{_{\textup{\scriptsize<#1>}}}#2)}
\newcommand{\mstart}[0]{\text{\textasciicircum{}}}
\newcommand{\mend}[0]{\mathdollar}
\newcommand{\wboundary}[0]{\escaped{\textup{b}}}
\newcommand{\wBoundary}[0]{\escaped{\textup{B}}}
\newcommand{\lookahead}[1]{({?}{=}\ #1)}
\newcommand{\lookbehind}[1]{({?}{\leq}\ #1)}
\newcommand{\neglookahead}[1]{({?}{\not=}\ #1)}
\newcommand{\neglookbehind}[1]{({?}{\not\leq}\ #1)}
\newcommand{\nbackref}[1]{\escaped{\textup{#1}}}
\newcommand{\backref}[1]{\escaped{\textup{#1}}}

\def\figcodesize{\footnotesize}

\author{Noé De Santo}
\email{noe.desanto@epfl.ch}
\orcid{0009-0006-5119-3895}

\author{Aurèle Barrière}
\email{aurele.barriere@epfl.ch}
\orcid{0000-0002-7297-2170}

\author{Clément Pit-Claudel}
\email{clement.pit-claudel@epfl.ch}
\orcid{0000-0002-1900-3901}

\affiliation{%
  \institution{EPFL}
  \city{Lausanne}
  \country{Switzerland}}

\setcopyright{rightsretained}
\acmDOI{10.1145/3674666}
\acmYear{2024}
\acmJournal{PACMPL}
\acmVolume{8}
\acmNumber{ICFP}
\acmArticle{270}
\acmMonth{8}
\acmSubmissionID{icfp24main-p109-p}
\received{2024-02-28}
\received[accepted]{2024-06-18}

\ccsdesc[500]{Software and its engineering~Semantics}

\keywords{ECMAScript, Regex, Mechanization, Coq}

\begin{abstract}
  We present an executable, proven-safe, faithful, and future-proof Coq mechanization of JavaScript regular expression (regex) matching, as specified by the latest published edition of ECMA-262 section 22.2.
  This is, to our knowledge, the first time that an industrial-strength regex language has been faithfully mechanized in an interactive theorem prover.
  We highlight interesting challenges that arose in the process (including issues of encoding, corner cases, and executability), and we document the steps that we took to ensure that the result is straightforwardly auditable and that our understanding of the specification aligns with existing implementations.

  We demonstrate the usability and versatility of the mechanization through a broad collection of analyses, case studies, and experiments: we prove that JavaScript regex matching always terminates and is safe (no assertion failures); we identify subtle corner cases that led to mistakes in previous publications; we verify an optimization extracted from a state-of-the-art regex engine; we show that some classic properties described in automata textbooks and used in derivatives-based matchers do not hold in JavaScript regexes; and we demonstrate that the cost of updating the mechanization to account for changes in the original specification is reasonably low.

  Our mechanization can be extracted to OCaml and JavaScript and linked with Unicode libraries to produce an executable regex engine that passes the relevant parts of the official Test262 conformance test suite.
\end{abstract}

\begin{document}
\title{A Coq Mechanization of JavaScript Regular Expression Semantics}
\maketitle

\section{Introduction}%
\label{sec:intro}

Like most modern programming languages, JavaScript natively offers a way to write and match regular expressions (or \emph{regexes}).
These regexes are a convenient tool for text processing --- so convenient in fact that a large proportion of JavaScript programs rely on them, including more than 30\% of JavaScript \texttt{npm} packages~\cite{redos_impact,lingua_franca}.
The semantics of JavaScript is described in an open standard, ECMAScript~\cite{ecma_262}, that specifies the behavior of the entire language with pseudocode.

The regex part of ECMAScript is surprisingly subtle and complex.
This complexity comes from the presence of extended regex features not found in traditional regular expressions (\emph{capture groups}, \emph{backreferences}, \emph{lookarounds}, \emph{character classes}, \emph{lazy quantifiers}, etc.; \cref{sec:ecma}), but also from some unconventional choices in the design of these features: even though other regex languages (like the ones in Python, Java, Go, Rust, Perl) share some of them, their semantics often differ in subtle ways~\cite{lingua_franca}.
For instance, JavaScript lookarounds are different from Java lookarounds, and JavaScript quantifiers and capture groups have unique semantics~\cite{linear_matching_js}.
To make matters worse, the ECMAScript regex standard is in continuous evolution: every recent revision of the standard contained significant additions or refactorings in the regex section, with dozens of individual edits every year;
this churn caused Mozilla to abandon the development of its own regular expression engine and switch to Irregexp, an engine developed as part of the V8 JavaScript implementation used in Google Chrome and NodeJS, to better keep up with new features~\cite{spidermonkey_irregexp}.

With all these features, JavaScript regex matching becomes significantly different from traditional regular expression recognition.  First, the addition of backreferences changes the matching complexity from linear to NP-hard~\cite{nphard}.  Second, the addition of capture groups changes the problem from recognition to segmentation: engines must not just compute whether a pattern matches a string, but also where the match occurred and which substrings were matched by each individual capture group (subexpressions in parentheses; this process is called \textit{submatch extraction}).
For instance, when matching the regex $\ch{a}(\ch{b}\star{})\ch{c}$ on the string \str{abbcd}, an engine should return that there is a match, but also that the regex matched the four first characters \str{abbc} and that the capture group is defined and set to the value \str{bb}.
Supporting capture groups is very convenient, but it introduces ambiguity in the matching problem: when there are several ways to match a regex on a string, the language semantics must specify which way has the highest priority.
In JavaScript, priority is given to the left branch of disjunctions.
For instance, matching the regex $\group{\ch{a} \disj{} \ch{a}\ch{b}}$ on the string \str{ab} returns a result that only matches the substring \str{a} (this is in line with PCRE greedy match semantics, but not with POSIX leftmost longest match semantics).
This disambiguation policy is particularly easy to implement --- and to specify --- using a backtracking algorithm.

For all the reasons above, the ECMAScript specification describes the JavaScript regex semantics with English pseudocode for a backtracking algorithm.  There are multiple issues with this approach.

First, the pseudocode is not executable.  It cannot be used directly to validate implementations of JavaScript, and it cannot be run to clarify ambiguities or validate proposed test cases.
As a result, bugs continue to be found even in implementations of long-standing features. For example, until recently, one of the engines included in V8 contained bugs because of a misunderstanding of the semantics of the Kleene star quantifier~\cite{linear_matching_js}.

Second, the length and style of the specification make it unsuitable for formal reasoning.
Mechanized semantics exist for substantial parts of the ECMAScript standard~\cite{jiset,jscert,lambdajs,aplas_js,kjs}, but none of them support JavaScript regexes.
As a result, previous work that tried to reason about JavaScript regexes had to redefine their own models of the semantics~\cite{regex_repair,expose}.
Unfortunately, of the two published models that we are aware of, one is incomplete, and we found mistakes in both (\cref{subsec:js_formalizations}).
To enable trustworthy reasoning about JavaScript regexes, one needs a faithful and auditable mechanized transcription of the ECMAScript pseudocode, without simplifications.

In that context, we present an \textbf{executable, proven-safe, faithful and future-proof Coq mechanization of JavaScript regexes}, as specified by the 2023 edition of ECMA-262 section 22.2.
By \emph{executable}, we mean that we can extract our mechanization either to OCaml or to JavaScript and run it.
By \emph{proven-safe}, we mean that we have a Coq proof that the compilation and matching process described in the specification terminates and that all assertions found in the specification are correct.
By \emph{faithful}, we mean that our mechanization follows the specification to the letter so as to be straightforwardly auditable (it is written in such a way that one can easily compare the pseudocode to the Coq definitions: we purposefully do not attempt to simplify the specification), and that it agrees with the spirit of the specification as understood by existing implementations.
Finally, by \emph{future-proof}, we mean that changes to the ECMAScript pseudocode require changes of comparable size to our mechanization, ensuring that it can be updated in the future as the standard evolves.

We claim the following contributions:
\begin{itemize}
\item \itemname{A shallow-embedded Coq mechanization} We show that ECMAScript regex semantics can be shallowly embedded in Coq, and we describe the encoding techniques required to represent exceptions, nonlocality, and possible nontermination in that style (\cref{sec:mechanization}).  We match the original specification as closely as possible: throughout the development, we interleave Coq code and English statements so that one can easily audit the mechanization. In total, we translated over 33 pages of pseudocode into Coq definitions.
\item \itemname{Safety and termination proofs} Using our mechanization, we prove the correctness of several non-trivial properties of the ECMAScript regex semantics, demonstrating that our mechanization enables formal reasoning about JavaScript regexes. In \cref{sec:proofs}, we prove that the specification always terminates and that it cannot crash. In the process, we surface and mechanize crucial invariants.
\item \itemname{A usable foundation for regex research} We show that our mechanization style is compatible with traditional program proofs and provides a usable formal foundation for regular expression compilation and optimization research by verifying a regex optimization excerpted from the Irregexp engine found in V8 (\cref{sec:strictly_nullable}).
\item \itemname{An executable ground-truth for JavaScript regexes} We show that our mechanization style is compatible with extraction to OCaml and compilation to JavaScript. We link the result with Unicode libraries, and show that we can plug it into a JavaScript engine in a way that passes all relevant standard tests, albeit slowly in some cases (\cref{subsec:extraction}).  To the best of our knowledge, this is the first time that a candidate reference interpreter has been proposed for JavaScript regexes.
\end{itemize}

We start by giving a presentation of JavaScript regexes, their semantics, and related work formalizing them in \cref{sec:ecma}.
We then describe our mechanization in \cref{sec:mechanization}, highlighting the translation challenges that arise and the encoding techniques that make a shallow embedding possible and scalable (an error monad for failing operations of the specification, zipper contexts for nonlocal operations, and a proven-correct fuel encoding for general recursion).
In \cref{sec:proofs}, we conjecture and formally verify a fundamental invariant of the ECMAScript regex specification, and we use it to prove that the specification always terminates and cannot fail.
Then, in \cref{sec:strictly_nullable}, we present the formal verification of a regex transformation performed in Irregexp.
Finally, in \cref{sec:evaluation}, we present evidence that our mechanization is indeed executable, proven-safe, faithful, and future-proof.
Specifically, we explain how to use extraction to generate two executable engines, one in OCaml and one in JavaScript. We show that the JavaScript engine can execute the Test262 official ECMAScript compliance test suite, and we evaluate on one example the effort required to extend our mechanization to support a new regex feature.

\section{Understanding JavaScript Regexes}%
\label{sec:ecma}

We begin this section with an overview of JavaScript regexes in \cref{subsec:features}.
We then describe their backtracking ECMAScript specification in \cref{subsec:matching}.
In \cref{subsec:invariant}, we give an informal explanation of a crucial property of all functions generated by the specification; we later mechanize this property to prove properties about the semantics (\cref{sec:proofs}).
Finally, in \cref{subsec:js_formalizations}, we present some problems that we uncovered in previous formalizations of JavaScript regexes and illustrate the semantic subtleties that led to them.

\subsection{Features of JavaScript Regexes}%
\label{subsec:features}

The ECMAScript standard for regexes includes not only standard regular expression features (concatenation, disjunction, simple characters and Kleene quantifiers like $\star{}$ or $\plus{}$), but also a number of extended features~\cite[Section 22.2]{ecma_262}.
\Cref{fig:ecma-features} summarizes the syntax of most of these features (we deviate from the plain-text JavaScript syntax in minor ways for readability purposes: for example, we write $\epsilon$ instead of nothing at all and $\lookbehind{r}$ instead of \texttt{(?<=r)}).

\begin{figure}
  \centering
  \begin{tabular}{lcc}
    Feature name & Explanation & Syntax \\\toprule
    Empty regex & & $\epsilon$ \\
    Character & & \ch{a}, \ch{b}, \ch{c}, \escaped{$n$}, \ldots \\
    Character classes & & \charclass{\ch{abc}}, \ncharclass{\ch{A}-\ch{Z}}, \dotc{}, \escaped{d}, \ldots \\
    Concatenation & & $r_1 r_2$ \\
    Disjunction (union) &  & $r_1 \disj{} r_2$ \\
    Greedy quantifiers & \ref{expl:quantifiers} & $r\star{}$, $r\plus{}$, $r\question{}$, $r\rrep{n}{n}$, \ldots \\
    Lazy quantifiers & \ref{expl:quantifiers} & $r\lstar{}$, $r\lplus{}$, $r\lquestion{}$, $r\lrrep{n}{n}$, \ldots \\
    Capturing groups & \ref{expl:groups} & $\group{r}$, $\group[\ensuremath{\mathit{id}}]{r}$ \\
    Non-capturing groups & \ref{expl:nc-groups} & $\pgroup{r}$ \\
    Backreferences & \ref{expl:backrefs} & $\escaped{p}$, $\nbackref{\ensuremath{\mathit{id}}}$ \\
    Lookarounds & \ref{expl:lookarounds} & $\lookahead{r}$, $\lookbehind{r}$, $\neglookahead{r}$, $\neglookbehind{r}$ \\
    Anchors & \ref{expl:anchors} & $\mstart{}$, $\mend{}$, $\wboundary{}$, $\wBoundary{}$
  \end{tabular}\\[0.4em]
  where $n$ is a nonnegative integer, $p$ is a positive integer, \ensuremath{\mathit{id}} is a named-group identifier (a string), $r$, $r_1$ and $r_2$ are regexes.
  \Description{}
  \caption{Features of the ECMAScript regex language.}%
  \label{fig:ecma-features}
\end{figure}

We now explain some of the extended features relevant to the remainder of this work, then contrast their semantics with those of the same features in other languages.

\begin{enumerate}[label=(\alph*)]

\item\label{expl:quantifiers}
  \itemname{Quantifiers}
  specify that a regex should be matched repeatedly.
  $r\rrep{n_1}{n_2}$ means that the regex $r$ should be matched between $n_1$ and $n_2$ times.
  If $n_2$ is not specified, then $r$ can be matched an unbounded number of times.
  By definition $r\star{} = r\rrep{0}{}$, $r\plus{} = r\rrep{1}{}$, and $r\question{} = r\rrep{0}{1}$.

  \emph{Greedy} quantifiers, like $\star{}$, $\plus{}$, and $\rrep{n_1}{n_2}$, give priority to matching the inner regex as many times as possible. \emph{Lazy} quantifiers, like $\lazystar{}$, $\lazyplus{}$, and $\lrrep{n_1}{n_2}$ give priority to matching the inner regex as few times as possible.
  For instance, $\ch{a}\plus{}$ matches the first three letters of the string \str{aaab} while $\ch{a}\lazyplus{}$ only matches the first one.
  To avoid infinite repetitions, the ECMAScript semantics forbids optional repetitions that match the empty string.
  For instance, when matching $()\plus{}$, the semantics allows only one iteration of the plus.
  The first iteration is allowed (in fact, mandated) because the plus has one minimal mandatory repetition, but all further iterations are disallowed: they are optional, the body $()$ of the quantifier matches only the empty string, and empty matches are not allowed for empty strings.

\item\label{expl:groups}
  \itemname{Capture groups}
  are used to record which part of the input string matched each paren-delimited sub-group of the original regex.
  When a regex matches a string, the engine must return not only the substring that was matched by the entire regex, but also all the substrings that were last matched by each capture group.
  For instance, consider the regex $\ch{a}\star{}\group{\ch{b}\star{}}\group{\ch{a}\star{}}$, with two capture groups, being matched on string \str{abbaaac}.
  The expected result is that the regex matches the substring \str{abbaaa}, that the first capture group $\group{\ch{b}\star{}}$ matches the substring \str{bb}, and that the second capture group $\group{\ch{a}\star{}}$ matches the substring \str{aaa}.
  Capture groups are numbered from 1 in order of appearance of their left parenthesis (this corresponds to a pre-order traversal of the AST).
  It is also possible to name capture groups (e.g.\ $\ch{a}\star{}\group[B]{\ch{b}\star{}}\group[A]{\ch{a}\star{}}$): this makes it possible to retrieve submatches by name and to refer to the group by name in backreferences.

\item\label{expl:nc-groups}
  \itemname{Non-capturing groups}
  act like parentheses do in mathematics: they override (or make explicit) operator priorities. For instance, the $\star{}$ quantifier is applied to $\ch{b}$ in $\ch{a}\ch{b}\star{}$ but to $\ch{a}\ch{b}$ in $\pgroup{\ch{a}\ch{b}}\star{}$.

\item\label{expl:backrefs}
  \itemname{Backreferences}
  match the text that was captured by an earlier group.
  For instance, the regex $\group[A]{\ch{a}\star{}}\group[B]{\ch{b}\star{}}\nbackref{A}$ matches strings of the form {\ch{a}$^n$\ch{b}$^m$\ch{a}$^n$} (two sequences of `\ch{a}'s of the same length separated by a sequence of `\ch{b}'s).
  Backreferences can also refer to unnamed groups by index.
  For instance, $(\ch{a}|\ch{b})\backref{1}$ matches \str{aa} or \str{bb} but not \str{ab}.
  Referring to a yet-undefined capture group (for instance for the regex $\nbackref{1}(\ch{a})$) is not an error: the backreference matches the empty string.

\item\label{expl:lookarounds}
  \itemname{Lookarounds}
  condition matching on the surrounding context: they allow matching to proceed only if the preceding or following text matches another regex.
  For instance, the regex $\neglookbehind{\ch{c}}\group[A]{\ch{a}\star{}}\group[B]{\ch{b}\star{}}\nbackref{A}$  matches the same strings {\ch{a}$^n$\ch{b}$^m$\ch{a}$^n$} as in the paragraph above, but only when not preceded by a `\ch{c}', because of the negative lookbehind $\neglookbehind{\ch{c}}$.  Similarly, with a positive lookbehind, $\lookbehind{\ch{b}}\group[A]{\ch{a}\star{}}\group[B]{\ch{b}\star{}}\nbackref{A}$ matches strings of the form {\ch{a}$^n$\ch{b}$^m$\ch{a}$^n$} but only when preceded by a `\ch{b}' (lookarounds are \emph{zero-width}, meaning that the initial `\ch{b}' is not part of the matched substring: the assertion only ensures that there is a `\ch{b}' before the match).
  Lookaheads instead condition the matching on the remainder of the string. For instance, $\ch{a}\lookahead{\ch{b}}$ will match the letter `\ch{a}' only when it is followed by a `\ch{b}'.
  Lookarounds include both lookaheads and lookbehinds, and can be either positive or negative.

  \item\label{expl:anchors}
  \itemname{Anchors}, sometimes also called \emph{boundary assertions}, are similar to lookarounds: they condition the match on surrounding characters. $\mstart{}$ and $\mend{}$ assert that there are no characters before and after them respectively. $\wboundary{}$ asserts that either the character before it or after is a word character, but the other is not. Word characters are all characters treated as equivalent to letters and numbers by the Unicode specification~\cite[Section 22.2.2.9.2]{ecma_262}.
  For instance, $\mstart{}\ch{a}$ does not find a match in the string \str{ba} since the `\ch{a}' is not at the beginning of the string.
\end{enumerate}

\paragraph{ECMAScript flags}
JavaScript regexes may be annotated with flags to configure the matching process.
Examples include the \inlinecode{i} flag to make the matching process case-insensitive, the \inlinecode{d} flag to compute submatch positions, or the \inlinecode{u} flag to enable Unicode mode (we give more details about this mode in \cref{subsec:extraction}).

\subsubsection{Comparison to Other Regex Languages}%
\label{subsubsec:comparison}
While similar features are found in most other regex languages, each language can make different choices of syntax and semantics~\cite{lingua_franca}.
These differences can be as simple as not using the same syntax: $\escaped{\textup{A}}$ is an anchor in languages like Python and PCRE, but simply matches the character `A' in JavaScript without the unicode flag.

But these differences can also be much more subtle.
For instance, capture groups inside quantifiers have unique semantics in JavaScript~\cite{linear_matching_js}.
Each time a quantifier is entered, the values of all capture groups inside it are reset to \inlinecode{undefined}~\cite[Section 22.2.2.3.1]{ecma_262}.
This is specific to JavaScript regexes, and it implies, for instance, that matching the regex $\pgroup{\group{\ch{a}} \disj{} \group{\ch{b}}}\star$ on the string \str{ab} defines group 2 ($\group{\ch{b}}$) but not group 1 ($\group{\ch{a}}$): the first iteration of the star defines group 1, but upon entering the second iteration of the loop all groups defined within the loop are reset, and after the second iteration only `\ch{b}' is set.
Similarly, quantifiers that can match the empty string behave differently in ECMAScript and in other regex languages~\cite{linear_matching_js}.

\subsection{ECMAScript Regex Matching}%
\label{subsec:matching}

To understand our mechanization, we now present the inner workings of the pseudocode algorithm that specifies ECMAScript regex matching.
The specification divides the process of matching a regex such as $\group{\ch{a}} \disj{} \ch{b}\ch{c}$ against an input string into four steps: parsing, validation (``early errors''), compilation, and execution:

\begin{enumerate}
\item \itemname{Parsing}~\cite[Section 22.2.1]{ecma_262}
  The textual representation of the regex is turned into an \emph{Abstract Syntax Tree} (AST).

  \item \itemname{Early errors}~\cite[Section 22.2.1.1]{ecma_262}
    Regexes with undesirable properties (called \emph{early errors}) are rejected.
    Examples include:
    \begin{itemize}
      \item $r\rrep{n_1}{n_2}$ when $n_1 > n_2$;
      \item $\group[G]{\ch{a}}\group[G]{\ch{b}}$, which defines the group G twice;
      \item $\nbackref{G}$, which has a dangling named backreference: the named group G is referred to but never defined.
    \end{itemize}

  \item \itemname{Compilation}~\cite[Section 22.2.2]{ecma_262}
    The AST of the regex is transformed into a matcher function which takes two arguments:
    \begin{commalist}
      \item an input string to search for a match
      \item and an index (a position in the string) to start the search at
    \end{commalist}.
    This function returns either \inlinecode{mismatch}\footnote{Named \inlinecode{failure} in the specification, but in this paper we prefer \inlinecode{mismatch}  to avoid confusion with assertion failures.}, indicating that no match was found, or a \inlinecode{MatchState}~\cite[22.2.2.1]{ecma_262} containing the input string, the index at which the match ended, and a description of the capture groups defined during the matching process.

  \item \itemname{Execution}~\cite[Section 22.2.5]{ecma_262}
    At a later point, when the user calls one of the JavaScript RegExp API functions (\texttt{match}, \texttt{matchAll}, \texttt{exec}, etc.), the generated matcher function is called through the \inlinejs{RegExp.prototype.exec} function.

\end{enumerate}

Our mechanization covers the last three: we delegate parsing to a preexisting regex parsing library for reasons that we detail in \cref{subsec:extraction}.
Our mechanization represents regex objects using an inductive type, with each constructor corresponding to a production rule of the ECMAScript regex-parsing grammar.

\subsubsection{ECMAScript Regex Compilation and Execution}

Perhaps surprisingly, the paper specification of ECMAScript regular-expression matching does not define an interpreter or an inductive matching relation.  Instead, the specification defines a compilation algorithm that transforms the parsed regex AST into a matching function, and separately explains how to execute the matching function on the input string.

\Cref{fig:types-sum} summarizes the different types used by the regex specification.
\inlinecode{MatchState} represents the internal state of the backtracking engine.
It contains the original input string, the current position in that string (called \inlinecode{endIndex}, an integer), as well as the capture indices of each group that has already been defined (\inlinecode{Captures}, an array of indices).
A \inlinecode{MatchResult} is either the \inlinecode{MatchState} obtained at the end of a match or the special value \inlinecode{mismatch} indicating that no match was found.
A \inlinecode{MatcherContinuation} is a function from a \inlinecode{MatchState} to a \inlinecode{MatchResult}.
Finally, a \inlinecode{Matcher} is a function taking a \inlinecode{MatchState} and a \inlinecode{MatcherContinuation} and returning a \inlinecode{MatchResult}.

\begin{figure}
\begin{tabular}{rcl}
  \inlinecode{MatchState} & \inlinecode{:=} & \inlinecode{(String} $\times$ \inlinecode{EndIndex} $\times$ \inlinecode{Captures)}\\
  \inlinecode{MatchResult} & \inlinecode{:=} & \inlinecode{MatchState} \, or \, \inlinecode{mismatch} \\
  \inlinecode{MatcherContinuation} & \inlinecode{:=} & \inlinecode{MatchState -> MatchResult} \\
  \inlinecode{Matcher} & \inlinecode{:=} & \inlinecode{MatchState -> MatcherContinuation -> MatchResult}
\end{tabular}%
\Description{}
\caption{Summary of the different types appearing in the compilation process.}%
\label{fig:types-sum}
\end{figure}

\begin{wrapfigure}{R}{7cm}
  \centering%
  \begin{forest}
    for tree={l sep=4mm, s sep=(3-level)*3mm}
    [{$r_1={}$Concatenation},name=1
      [{$r_2={}$Disjunction}, name=2
        [{$r_3={}$Concatenation}, name=3
          [{$r_4=\ch{a}$}, name=4]
          [{$r_5=\ch{b}$}, name=5]]
        [{$r_6=\dotc{}$}, name=6]]
      [{$r_7=\ch{b}$}, name=7]
    ]
  \end{forest}
  \caption{Abstract syntax tree of $r_1 = \pgroup{\pgroup{\ch{a}\ch{b}} \disj{} \dotc{}}\ch{b}$\vspace{-\baselineskip}}%
  \label{fig:matching-regex}\label{fig:regex-ast}
\end{wrapfigure}

The compilation is done by the function \inlinecode{compilePattern}~\cite[22.2.2.2]{ecma_262}, with most of the work happening in the function \inlinecode{compileSubPattern}~\cite[22.2.2.3]{ecma_262}.
The \inlinecode{Matcher}'s first argument is its initial state.
The second argument is a continuation function used to continue the matching if this matcher succeeds (initially, this continuation is the identity function).
During the execution of the matcher functions, this continuation function is used to encode regex concatenation.

Let us illustrate the backtracking behavior with an example.
Consider matching the string \str{ab} with the regex $r_1 = \pgroup{\pgroup{\ch{a}\ch{b}} \disj{} \dotc{}}\ch{b} $, whose AST is pictured in \cref{fig:matching-regex}.
The dot $\dotc$ is a regex construct matching any single character.
We write $m_i$ for the matcher compiled from the regex $r_i$.
A trace of the execution of the semantics is shown in \cref{fig:matching-progress}; we explain it here in more details.

\medskip
\begin{description}
  \item \itemname{Step 0}
    First, $m_1$ is called with a state \inlinecode{x0} at index 0 and a trivial continuation \inlinecode{c0} (the identity).

  \item \itemname{Step 1}
    $r1$ is a concatenation: as such, its compiled matcher first creates a new continuation \inlinecode{c7}.
    This continuation encodes the matching of the right side of the concatenation ($r_7 = \ch{b}$).
    It uses the matcher of $r_7$ (\inlinecode{m7}) and the initial continuation \inlinecode{c0} it was called with.
    With this new definition, \inlinecode{m1} calls the matcher of its left-hand side (\inlinecode{m2}) with the new continuation \inlinecode{c7}.

  \item \itemname{Step 2}
    \inlinecode{m2} is a disjunction.
    This is called a \textit{choice point}: the backtracking pseudocode must try the first alternative, and later maybe backtrack to try the other one.
    \inlinecode{m2} first calls the matcher of the left alternative (\inlinecode{m3}) with continuation \inlinecode{c7} to try to match $r_3 = \ch{a}\ch{b}$ followed by $r_7 = \ch{b}$.

  \item \itemname{Step 3}
    \inlinecode{m3} is again a concatenation, so \inlinecode{m4} is called with a new continuation (\inlinecode{c5}), as in step 1.

  \item \itemname{Steps 4--5}
    \inlinecode{m4} tries to consume an `\ch{a}' and succeeds. It advances the state, yielding \inlinecode{x1} at position 1, and calls its continuation \inlinecode{c5} with it.
    This continuation in turn invokes \inlinecode{m5 x1 c7}.

  \item  \itemname{Steps 6--7}
    \inlinecode{m5} tries to consume a `\ch{b}' and succeeds. It advances the state, yielding \inlinecode{x2} at position 2, and calls its continuation \inlinecode{c7}, which in turn invokes \inlinecode{m7 x2 c0}.
    At this point, the engine successfully matched \str{ab} using the left-hand alternative of the disjunction. But now, it must match $r_7 = \ch{b}$.

  \item \itemname{Step 8}
    \inlinecode{m7} tries to consume a `\ch{b}' but fails since the end of the input string has been reached.
    It hence returns \inlinecode{mismatch}.

  \item \itemname{Step 9: backtracking}
   Since the call to \inlinecode{m3 x0 c7} returned \inlinecode{mismatch}, we return to the call of \inlinecode{m2}.
   It now attempts a match using \inlinecode{m6}, which corresponds to $r_6$, the right-hand alternative.

  \item \itemname{Steps 10-13}
    The process continues as before for a few steps. At the end, the state \inlinecode{x4} is returned, which indicates that a match was found.
\end{description}
\medskip

The recursive structure of the generated matchers makes it so that their call stack acts as an implicit backtracking stack: every time a matcher has multiple ways to match a regex (for instance for a disjunction or a quantifier), it tries them one by one until one succeeds.  In contrast, when there is only one way to match, the matcher calls its continuation in tail position.

\begin{figure}
  \centering
  \begin{tabular}{p{2.5em}p{10em}p{10em}cl}
    Step & Call stack & States and continuations & TC\; & Comment \\ \toprule
    0 & \inlinecode{[]} & \inlinecode{x0 :=} ab (position 0) & & Initialization \\
    & & \inlinecode{c0 := fun s => s} & & \\
    1 & \inlinecode{[m1 x0 c0]} & \inlinecode{c7 := fun s => m7 s c0} & \cmark & Concatenation \\
    2 & \inlinecode{[m2 x0 c7]} & & & Choice point (disjunction) \\
    3 & \inlinecode{[m3 x0 c7, m2 x0 c7]} & \inlinecode{c5 := fun s => m5 s c7} & \cmark & Concatenation \\
    4 & \inlinecode{[m4 x0 c5, m2 x0 c7]} & \inlinecode{x1 :=} \underline{a}b (position 1) & \cmark & Consumes \str{a} \\
    5 & \inlinecode{[c5 x1, m2 x0 c7]} & & \cmark & \\
    6 & \inlinecode{[m5 x1 c7, m2 x0 c7]} & \inlinecode{x2 :=} \underline{ab} (position 2) & \cmark & Consumes \str{b} \\
    7 & \inlinecode{[c2 x2, m2 x0 c7]} & & \cmark & \\
    8 & \inlinecode{[m7 x2 c0, m2 x0 c7]} & & & Cannot consume \str{b} \\
    \midrule \multicolumn{5}{c}{Match failed $\implies{}$ Backtrack to last choice point} \\ \midrule
    9 & \inlinecode{[m2 x0 c7]} & & \cmark & \\
    10 & \inlinecode{[m6 x0 c7]} & & \cmark & \\
    11 & \inlinecode{[c7 x3]} & \inlinecode{x3 :=} \underline{a}b (position 1) & \cmark & Consumes \str{a} \\
    12 & \inlinecode{[m7 x3 c0]} & \inlinecode{x4 :=} \underline{ab} (position 2) & \cmark & Consumes \str{b} \\
    13 & \inlinecode{[c0 x4]} & & & Returns x4
  \end{tabular}
  \\[0.4em]
  A check mark in the ``TC'' column indicates that a call in tail position is performed.\\
  The underlined part of a \inlinecode{MatchState} indicates the characters that have already been consumed.%
  \Description{}
  \caption{Matching process for $r_1$ on input string \str{ab}.}%
  \label{fig:matching-progress}
\end{figure}

\subsection{Matcher Invariant}%
\label{subsec:invariant}

The details given above are enough to understand a key invariant of matcher functions generated by the specification, which we conjectured and proved in the course of our mechanization effort.

\bigskip

\paragraph{\textup{\textbf{Matcher Invariant:}}} any call to a \inlinecode{Matcher} function with a state \inlinecode{x} and a continuation \inlinecode{c} either returns a mismatch, or calls its continuation \inlinecode{c} on another state \inlinecode{y} such that the \inlinecode{endIndex} of \inlinecode{y} is greater than or equal to the one of \inlinecode{x} and that both \inlinecode{x} and \inlinecode{y} have the same input string.\footnote{The direction of the inequality is reversed for lookbehinds, since they traverse the string backwards. We omit this case for now and generalize the invariant in \cref{subsec:generalize_progress}.}
Intuitively, this corresponds to the fact that the strings matched by the two halves of a concatenation do not overlap.

This invariant is not only critical to understand the compilation process and the generated matchers, but also to formally reason about them, both when proving rewrites and when proving semantic properties.
We formalize and prove a generalization of it in \cref{sec:proofs}.

\subsection{Semantic Subtleties Missed by Previous JavaScript Regex Formalizations}%
\label{subsec:js_formalizations}

Ours is the first mechanization of the ECMAScript regex standard, but not the first attempt to reason about JavaScript regexes formally: paper-based models of ECMAScript regex matching have appeared in multiple past publications~\cite{expose,regex_repair}.
These models are typically incomplete (omitting complex features such as lookbehinds~\cite{expose}).  More worryingly, our efforts to faithfully mechanize the specification revealed that these models were also incorrect.  In this section, we illustrate some of the most subtle issues.

\paragraph{Rewriting greedy and lazy quantifiers}
The models presented in~\citet{expose} and~\cite{regex_repair} incorrectly handle greedy and lazy question mark operators, $\question{}$ and $\lquestion{}$.  They assume that $r\question{}$ can be equivalently rewritten to $\pgroup{r \disj{}\epsilon} $, and that $r\lquestion{}$ can be rewritten to the regex $\pgroup{\epsilon\disj{}r}$, allowing for a smaller and cleaner language without $\question{}$ and $\lquestion{}$.

Unfortunately, both of these transformations are incorrect in general.  In particular, a problem occurs when $r$ can match the empty string while defining capture groups.

For greedy question marks like $r\question{}$, consider the regex $r = ()$, and compare the semantics of $r\question{}$ and $\pgroup{r\disj{}\epsilon}$ on the empty string.
In the $r\question{}$ case, the specification (\cref{subsec:features}) allows at most one optional iteration of $r$, but this iteration may not match the empty string.  Hence, the correct behavior is to not match $r$ at all (the quantifier $\question{}$ does 0 iterations).
The final result simply matches the empty string and does not define any capture groups.
In contrast, $\pgroup{r\disj{}\epsilon}$ allows $r$ to match the empty string.  Hence, the final result in this case does define the first capture group (the one contained in $r$).

For lazy question marks like $r\lquestion{}$, consider the regex
$r = \lookahead{\group{\ch{a}}}\lquestion{}\ch{a}\ch{b}\nbackref{1}\ch{c}$ being matched on string \str{abac}.
A compliant engine would not find a match.  Matching would proceed as follows:
the engine would first skip the lazy quantifier (doing so, the first capture group would not be defined), then consume \str{ab}, then match the empty string with the backreference (backreferences to undefined groups match the empty string), and finally attempt to match \str{c}.  Because the input is \str{abac}, the last step would fail, and the engine would backtrack.
Backtracking would lead the engine to attempt to match the body of the lazy question mark and define the capture group $\group{\ch{a}}$.
Unfortunately, because lookaheads are zero-width, this would lead to an empty iteration of the quantifier, which the specification disallows.  Having exhausted all backtracking opportunities, a correct engine would then return a match failure.

Rewriting the regex to $\pgroup{\epsilon\disj{}\lookahead{\group{\ch{a}}}}\ch{a}\ch{b}\nbackref{1}\ch{c}$ would not preserve this subtle behavior: unlike the original regex, this one does match \str{abac}.
Taking the empty left branch $\epsilon$ would not find a match, since it would not define the capture group.
The semantics then backtracks and takes the right branch of the disjunction (unlike previously, this would now be allowed).
Doing so, the engine would go into the lookahead and set the first capture group to \str{a}.  With that, matching as a whole would (incorrectly) succeed.

\paragraph{Incorrect claims about infinite loops}
It is sometimes mentioned~\cite{regex_to_transducers} that JavaScript and RE2 have the same way to prevent infinite looping in a star matching the empty string, which is not true and can lead to different match results, for instance for the regex $\pgroup{\ch{a}\question{}\ch{b}\lquestion{}}\star{}$ on string \str{ab}~\cite{linear_matching_js}.

\paragraph{Leveraging mechanized specifications to validate simplified semantics} Although we quickly found a counterexample for the first transformation, we initially believed that the second one was correct.
We then tried to formally verify it using our mechanization, and this attempt led us to the counterexample above.

The official ECMAScript regex standard is verbose and complex: it is no surprise that previously published models of JavaScript regular expressions attempted to capture regex semantics in a more succinct and practical style~\cite{regex_repair, expose}.  We too would prefer to reason about regexes in such a style, but it cannot be at the cost of correctness: to be trustworthy, formalizations of JavaScript regexes must follow the ECMAScript standard to the letter and avoid semantically incorrect simplifications.

We hope that our mechanization will provide foundations to \emph{prove}, rather than posit, the correctness of more succinct, readable, and practical specifications for JavaScript regex semantics.

\section{A Shallow-Embedded Coq Mechanization of JavaScript Regexes}%
\label{sec:mechanization}

In this section, we present our Coq mechanization of the ECMAScript standard for JavaScript regexes.
We describe a collection of encoding and translation techniques that make a manual faithful shallow embedding possible in Coq, and scalable to the size of the specification (about 33 printed pages).
Our main goal was to produce a mechanization as faithful and close as possible to the original pseudocode specification, and we use two key ingredients to ensure that.

\paragraph{Shallow embedding} First, we use a shallow embedding, where each function and type that appears in the pseudocode is directly translated to a similar function or type in the mechanization.
This is the most natural and straightforward way to have an equivalent mechanization.\footnote{Historical comments made by one of the original authors of the ECMAScript regex specification~\cite{tc39notes} suggest that the backtracking pseudocode may have been derived from a lambda-calculus-based model.  In that sense, mechanizing the specification in Coq may be returning it to its roots.}

\begin{figure}
\begin{subfigure}{\textwidth}
\includegraphics[width=\textwidth]{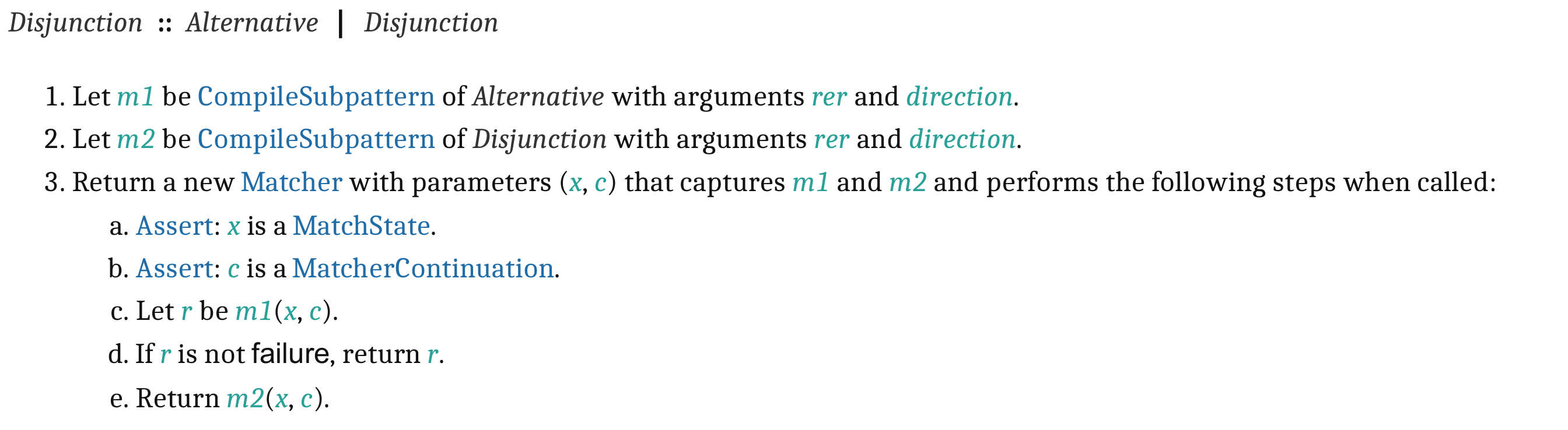}
\Description{}
\caption{A sample of the specification~\cite[Section 22.2.2.3]{ecma_262}.}%
\end{subfigure}
\begin{subfigure}{\textwidth}
\begin{lstlisting}[language=Coq,breaklines=true,basicstyle=\figcodesize]
(** >> Disjunction :: Alternative | Disjunction <<*)
| Disjunction r1 r2 =>
  (*>> 1. Let m1 be CompileSubpattern of Alternative with arguments rer and direction. <<*)
  let! m1 =<< compileSubPattern r1 (Disjunction_left r2 :: ctx) rer direction in
  (*>> 2. Let m2 be CompileSubpattern of Disjunction with arguments rer and direction. <<*)
  let! m2 =<< compileSubPattern r2 (Disjunction_right r1 :: ctx) rer direction in
  (*>> 3. Return a new Matcher with parameters (x, c) that captures m1 and m2 and performs the following steps when called: <<*)
  (fun (x: MatchState) (c: MatcherContinuation) =>
    (*>> a. Assert: x is a MatchState. <<*)
    (*>> b. Assert: c is a MatcherContinuation. <<*)
    (*>> c. Let r be m1(x, c). <<*)
    let! r =<< m1 x c in
    (*>> d. If r is not failure, return r. <<*)
    if r is not failure then r else
    (*>> e. Return m2(x, c). <<*)
    m2 x c): Matcher
\end{lstlisting}
\caption{Our mechanization of the sample above.}
\end{subfigure}%
\caption{Comparison of the paper specification with our mechanization.}%
\label{fig:sample-impl}
\end{figure}

\paragraph{Line-by-line auditability} Second, we interleave each line of our Coq mechanization with English statements taken directly from the corresponding pseudocode in the ECMAScript specification.
We see two main benefits to this style.
First, it eases the burden of auditing our mechanization to check that it performs the same operations as the ECMAScript standard.
\Cref{fig:sample-impl} shows an example of how close the two are once we define adequate notations and encodings.
Second, it gives us a measure of future-proofness: if the ECMAScript standard evolves (for instance, by adding new regex features), this style ensures that changes to our mechanization remain commensurate to the changes in the original specification.

We start by presenting our solution to encode the potentially failing operations used by the ECMAScript pseudocode as a pure Coq implementation in \cref{subsec:imperative}.
Some operations are even \emph{nonlocal}: they refer to the parent or siblings of the current AST node: we propose an adequate zipper encoding in \cref{subsec:nonlocal}.
Another issue when defining Coq functions is that Coq enforces termination. In \cref{subsec:termination}, we show how we encoded the loops and recursive functions used in the standard with a fuel parameter.

Our monadic encoding, zipper, and fuel parameter are simple and time-tested solutions that ensure a low barrier of entry to our Coq mechanization.
We converged on them after trying various other solutions, including encoding the termination and absence of failures with dependent types, and finally choosing the most straightforward encodings to update and reuse.

\subsection{Faithfully Encoding Potentially Failing Pseudocode in Coq}%
\label{subsec:imperative}

In various places, the specification makes use of language features that are not available in Gallina, the programming language of Coq.
The most common ones are \emph{assertions} and other potentially \emph{failing} operations.
Assertions are used extensively in the specification, but there is no such feature in Gallina.
One possible solution would be to ignore assertions entirely.
In fact, the ECMAScript standard itself states that:
\begin{quotation}
``Such assertions are used to make explicit algorithmic invariants that would otherwise be implicit. Such assertions add no additional semantic requirements and hence need not be checked by an implementation. They are used simply to clarify algorithms.''~\cite[Section 5.2]{ecma_262}.
\end{quotation}

Instead of ignoring assertions, our approach consists in encoding them within our Coq mechanization, using the Coq type system and an error monad.
Assertions can be split into two categories:
\begin{itemize}
\item Type assertions: assertions of the form \inlinespec{Assert: x is a MatchState} (e.g. \cref{fig:sample-impl} step \sstep{3.a}). For these, we use the type system of Gallina to ensure that the assertion holds.
\item Other assertions: assertions such as \inlinespec{Assert: xe <= ye}. For these, we use an error monad.
\end{itemize}

This approach has multiple benefits.
First, we keep the mechanization close to the pseudocode specification.
Second, it allows us to encode other potentially failing operations of the specification, like accessing the arrays holding the capture information in each \inlinecode{MatchState}.
Third, keeping assertions helps understand and reason formally about JavaScript regex behavior.

On the other hand, by adding assertions, we introduce potential failures where the paper specification claimed that assertions add no additional semantic requirements.  Consequently, we go even further and prove, in Coq (\cref{subsec:property_proofs}), that all assertions always hold and that other potentially failing operations never fail.

\paragraph{Proving safety with an error monad}
\label{sec:error-monad}

\begin{figure}
\begin{minipage}{.5\textwidth}
\begin{lstlisting}[language=Coq,breaklines=true,basicstyle=\figcodesize]
Inductive Result (S: Type) :=
 | Success (s: S)
 | Failure
 | OutOfFuel.

Definition bind m f :=
  match m with
  | Success v => f v
  | Failure   => Failure
  | OutOfFuel => OutOfFuel
  end.
\end{lstlisting}
\end{minipage}%
\begin{minipage}{.5\textwidth}
\begin{lstlisting}[language=Coq,breaklines=true,basicstyle=\figcodesize]
Notation "'let!' r '=<<' y 'in' z" :=
  (bind y (fun r => z)).

Notation "'assert!' b ';' z" :=
  (if (negb b) then Failure else z).

(* Non-exhaustive match *)
Notation "'destruct!' r '<-' y 'in' z" :=
  (match y with
   | r => z
   | _ => Failure
   end).
\end{lstlisting}%
\end{minipage}
\Description{}
\caption{A simple error monad and its notations.}%
\label{fig:error-monad}
\end{figure}

A simple error monad, as shown on \cref{fig:error-monad}, can faithfully encode assertions and other potentially failing operations of the specification.
\inlinecode{Failure} represents assertion failures and other kinds of failing operations (e.g.\ out-of-bounds array accesses), and \inlinecode{OutOfFuel} is a special case of failure that we use when encoding potentially non-terminating functions (see \cref{subsec:termination}).
The \inlinecode{bind} notation chains operations together and achieves a result that is visually very close to the pseudocode specification, as demonstrated in \cref{fig:sample-impl}.
Note that the specification also sometimes uses non-exhaustive pattern matching (for instance, assuming that some match result is not a mismatch), which can be encoded with our \inlinecode{destruct!} notation (we later prove that the missing branches are indeed never taken).

Every function of the specification is then wrapped in the error monad, from the compilation phase (compilation can fail, for instance, if a backreference refers to an undefined group) to the execution phase (the generated matcher functions can fail, for instance, if they access the capture array of a \inlinecode{MatchState} out of bounds).

In summary, all assertions of the ECMAScript standard are encoded either with type annotations or within the error monad, and all functions that return a value of type \inlinecode{T} while including potentially failing operations are changed to instead return a value of type \inlinecode{Result T}.

\begin{wrapfigure}{l}{6cm}
\centering
\includegraphics[height=1.5cm]{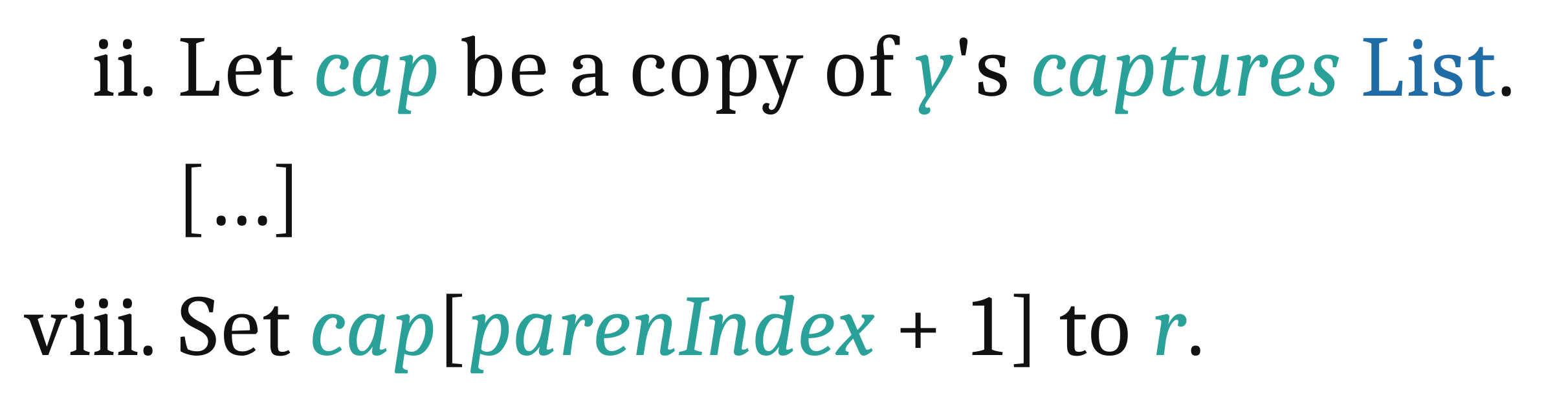}
\Description{}
\caption{Modifying a list in the compilation of groups~\cite[22.2.2.7]{ecma_262}.}%
\label{fig:list-mutation}%
\end{wrapfigure}

\paragraph{Avoiding mutation by recognizing specific usage patterns}
Besides potential failures, the specification uses another imperative feature: mutable lists.
In general, we would need a special encoding to handle mutable lists in a Coq mechanization.
For instance, we could augment our monadic encoding with a state monad to hold the current value of each mutable list.
However, the regex specification only uses mutable lists in very restricted cases, and they are always copied right before being modified.
For example, when compiling capture groups~\cite[Section 22.2.2.6]{ecma_262}, a list is mutated at step \sstep{3.c.viii}, but this list is a local copy which was done at step \sstep{3.c.ii}. These two instructions are shown on \cref{fig:list-mutation}.
Given this, we chose to only use functional immutable lists in our Coq mechanization.
This keeps the mechanization simple to write and reason about while following the spirit of the standard.

\subsection{Encoding Nonlocal Operations with a Zipper Context}%
\label{subsec:nonlocal}

\begin{figure}
\includegraphics[width=\textwidth]{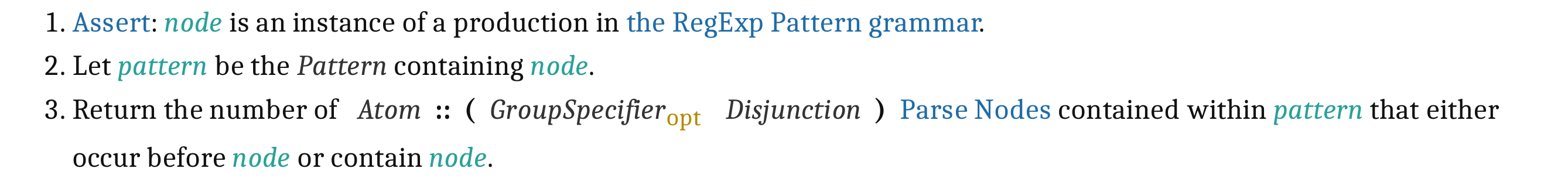}%
\Description{}
\caption{Specification of the \inlinecode{CountLeftCapturingParensBefore} function~\cite[22.2.1.3]{ecma_262}}%
\label{fig:high-level-spec}
\end{figure}

An implicit feature of the AST representation of regexes used by the specification is that it is possible to navigate the AST upward, which in particular allows it to retrieve the \emph{root} of the regex AST.\@
This is used in a few functions of the specification.
For example, as mentioned earlier, the index of a group is determined by counting the number of left parentheses appearing before it in the whole AST.\@ This is done by the \inlinecode{CountLeftCapturingParensBefore} function~\cite[22.2.1.3]{ecma_262}, shown in \cref{fig:high-level-spec}.
Consider the regex $r = \pgroup{(\ch{a})\disj{}(\ch{a})}\ch{b}$ defined in \cref{fig:non-loc-regex}.
\inlinecode{CountLeftCapturingParensBefore} should return 0 when called on sub-regex $r_1$ (the first capture group), as no parentheses appear above or to the left of it. When called on $r_2$  (the second capture group), the function should return 1 as there is one opening parenthesis on its left, namely the one in $r_1$.
This is one of the few functions in the ECMAScript regex pseudocode that are specified with such high-level descriptions, and without precise pseudocode instructions for each step.

This effectively means that regexes should include some form of context to relate them to the full regex they belong to, so that this function becomes feasible to implement.
We model this pattern using zippers~\cite{zipper}.
The zipper holds both the current AST node and its \emph{context}, allowing reconstruction of the root regex.
Additionally, comparing the context of two regexes allows to distinguish syntactically identical sub-regexes that appear at different locations in the AST.\@
Going back to the regex $r$, its child $r_2$ is represented by a pair of the AST for $\group{\ch{a}}$ and a context describing that this sub-regex is located on the left of the top-level concatenation, and then on the right of the disjunction: the context acts as a description of the path from the child to the root of the regex.
Additionally, since we also want to be able to reconstruct the root regex, this path is augmented with all the data necessary to perform the reconstruction: contexts can be implemented with a linked list of incomplete AST nodes.
These contexts are depicted on the right of \cref{fig:non-loc-regex}, where $\square$ shows the incomplete parts of each sub-regex in their contexts.
During compilation, as we traverse the AST of a regex, we manually maintain the associated context of each visited sub-regex.

\begin{figure}\centering
  \begin{subfigure}[t]{0.35\linewidth}\centering
    \begin{forest}
      [{$r\ \textup{(root)}={}$Concatenation}, name=root
        [Disjunction
            [{$r_1 ={}$Group}, name=g1 [\ch{a}]]
            [{$r_2 ={}$Group}, name=g2 [\ch{a}]]]
        [\ch{b}]
      ]
    \end{forest}
    \caption{A regex $r$.}
  \end{subfigure}
  \hfill
  \begin{subfigure}[t]{0.3\linewidth}\centering
    \begin{forest}
      for tree={text=gray, edge=gray},
      [$\square \ch{b}$, name=root
        [$\square | (\ch{a}) $
            [Group, name=g1,for tree={text=black}, for descendants={edge=black}, [\ch{a}]]]]
      \node[left = 0.5cm of g2] (g1_label) {focus};
      \draw[->] (g1_label) to[out = east, in = west] (g1);
    \end{forest}
    \caption{The sub-regex $r_1$ and its\\context.}
  \end{subfigure}
  \hfill
  \begin{subfigure}[t]{0.3\linewidth}\centering
    \begin{forest}
      for tree={text=gray, edge=gray},
      [$\square \ch{b}$, name=root
        [$(\ch{a}) | \square$
            [Group, name=g2,for tree={text=black}, for descendants={edge=black}, [\ch{a}]]]]
      \node[right = 0.5cm of g2] (g2_label) {focus};
      \draw[->] (g2_label) to[out = west, in = east] (g2);
    \end{forest}
    \caption{The sub-regex $r_2$ and its\\context.}
  \end{subfigure}
  \Description{}
  \caption{A regex and its contextualized sub-regexes. Contexts are represented in gray.}%
  \label{fig:non-loc-regex}
\end{figure}

\subsection{Handling Arbitrary Recursion}%
\label{subsec:termination}

The termination of most functions in the specification follows directly from the fact that these functions are structurally recursive on a regex.
There are, however, notable exceptions: the \inlinecode{RepeatMatcher} function~\cite[22.2.2.3.1]{ecma_262} and some functions of sections 22.2.6 and 22.2.7~\cite{ecma_262}. Since these are not structurally recursive, they are not accepted as-is by Coq, because it cannot ensure their termination.

There are multiple ways to encode non-terminating functions in Coq developments, for instance using coinductive definitions.
We opted for a \emph{fuel}-based solution:
we added an extra fuel argument of type \inlinecode{nat} to these functions. If the function is called with a positive amount of fuel, then any subsequent recursive call is handed one unit less fuel; if the fuel is zero, then the function fails and returns \inlinecode{OutOfFuel}.
From Coq's point of view, the function is then structurally recursive on this extra argument, and hence acceptable.
Of course, this raises the question of whether there always exists an amount of fuel such that \inlinecode{OutOfFuel} is never returned.
In \cref{subsec:property_proofs}, we provide such a bound as a function of the regex and input-string sizes, and formally prove in Coq that this amount of fuel is sufficient to rule out all \inlinecode{OutOfFuel} failures.
Given that termination can be proved, we find this to be a pleasant way to encode such non-terminating functions (alternative encodings such as coinductive functions are often perceived as much harder to reason about, and might force us to deviate further from the original pseudocode).

\section{Deriving and Mechanizing New Semantic Properties: Validating the ECMAScript Specification}%
\label{sec:proofs}

The ECMAScript specification itself could be incorrect, in the sense that its pseudocode could contain mistakes: for instance, it could produce functions that crash or do not terminate.
In this section, we show that one crucial advantage of our mechanization is that we can formally verify semantic properties about the ECMAScript regex specification.
In particular, we prove that, for any regex and any input string, the specification never crashes (e.g.\ assertions never fail and arrays are never accessed out of bounds) and always terminates.
Although these are natural properties that one could expect from any regex specification, they can be quite difficult to infer just by looking at the pseudocode.
With our mechanization, we can prove these properties in Coq, and as a result validate the correctness of the ECMAScript specification.

We originally proved these two properties separately.
However, we later realized that both properties are direct consequences of the stronger matcher invariant that we informally described in \cref{subsec:invariant}.
We first present our mechanized matcher invariant and describe how it can be used to prove termination and absence of failures.
We then present interesting cases of the proof that the matcher invariant holds for any matcher generated by the specification.
Finally, in \cref{subsec:generalize_progress}, we explain how to generalize the simplified notion of progress used in this paper to the one found in the mechanization, which accounts for backwards input traversals performed by lookbehinds.

\subsection{Mechanizing the Matcher Invariant}%
\label{subseq:coq_inv}
\begin{wrapfigure}{L}{6cm}
\begin{lstlisting}[language=Coq,basicstyle=\figcodesize]
Definition matcher_invariant (m: Matcher) :=
(* For any valid state x, continuation c *)
forall x c, Valid x ->
  (* either there is no match or *)
  (m x c = Success mismatch) \/
  (* m produced a valid state y which *)
  (exists y, valid y /\
    (* made progress with respect to x *)
    x LEQSLANT y /\
    (* and was passed to c *)
    c y = m x c).
\end{lstlisting}%
\caption{Mechanized Matcher Invariant.}%
\label{fig:matcher_invariant}
\end{wrapfigure}
\newcommand{\progress}{\leqslant}
As explained in \cref{subsec:matching}, the matcher functions and the continuations compiled by the ECMAScript standard follow a particular control-flow, where each matcher function ends in a mismatch or in a call to its continuation.
This call to the continuation is performed on a matcher state whose \inlinecode{endIndex} is greater than or equal to that of the state given as argument to the matcher, meaning that the matcher may have consumed some characters in the string before calling its continuation.
In other words, every result of a matcher function call is either a mismatch or the result of calling its continuation after having progressed in the string.
As it turns out, this invariant can be used to prove the termination and the absence of failures in the specification.

Once again, for presentation purposes we make the assumption that regexes only ever traverse the string forward, from its beginning to its end.
This assumption is not true for lookbehinds (both positive and negative), which traverse the string backwards.
We discuss how our approach can be generalized to both directions in \cref{subsec:generalize_progress}.

We first define the invariant in Coq as depicted in \cref{fig:matcher_invariant}.
This invariant also captures the fact that matcher functions are only ever called on \emph{valid} states, i.e.\ states whose \inlinecode{endIndex} is a valid index of \inlinecode{input}.
To encode this notion of progress in the string, we define a relation between match states, $\progress$, such that $x \progress y$ when \inlinecode{x} and \inlinecode{y} have the same input string and \inlinecode{endIndex(x)} $\leq$ \inlinecode{endIndex(y)}.

\paragraph{Proving the matcher invariant}
With this definition, it is possible to prove that all matchers generated by compiling a regex satisfy the invariant.
The corresponding theorem, which we proved in Coq, is shown in \cref{fig:invariant_preserved}.
We explain the main challenges of the proof in \cref{subsec:property_proofs}.

\begin{wrapfigure}{l}{6cm}
\begin{lstlisting}[language=Coq,basicstyle=\figcodesize]
Theorem compiled_regex_invariant :
  (* For any compiled matcher m  *)
  forall r m, compileSubPattern r = Success m ->
    earlyErrors r = OK ->
    (* the matcher invariant holds. *)
    matcher_invariant m.
\end{lstlisting}%
\caption{Proving the Matcher Invariant.}%
\label{fig:invariant_preserved}
\end{wrapfigure}

\paragraph{ECMAScript regex properties}
Two important non-trivial properties of the ECMAScript regex semantics follow from the theorem of \cref{fig:invariant_preserved}:
termination (functions generated by the specification always \emph{terminate}, i.e.\ they never run out of fuel),
and safety (the functions \emph{cannot fail}, e.g.\ assertions always hold, and arrays and strings are never accessed out-of-bounds).

We state and prove the two theorems of \cref{fig:coq_properties}, where \inlinecode{init_state} is an initial matcher state containing the string to match and starting at index 0, and \inlinecode{identity_cont} is the continuation that always returns a success.
The two proofs are immediate consequences of the \inlinecode{compiled_regex_invariant} theorem by contradiction:
if the call to a matcher resulted in a lack of fuel or a failure, then because this result is not a \inlinecode{mismatch}, the continuation \inlinecode{identity_cont} itself would result in a lack of fuel or a failure, which is a contradiction.

Both termination and absence of failures are properties that one would expect about the specification, but that one cannot check formally without a mechanization.
These proofs demonstrate that our mechanization enables formal reasoning about the JavaScript regex semantics in a proof assistant.

\begin{figure}
\begin{minipage}{.5\textwidth}
\begin{lstlisting}[language=Coq,basicstyle=\figcodesize]
Theorem termination :
  (* For any regex r, string s *)
  forall r m s, compileSubPattern r = Success m ->
    earlyErrors r = OK ->
    (* the matcher cannot run out of fuel. *)
    m (init_state s) (identity_cont) <> OutOfFuel.
    \end{lstlisting}
    \end{minipage}%
    \begin{minipage}{.5\textwidth}
\begin{lstlisting}[language=Coq,basicstyle=\figcodesize]
Theorem no_failure :
  (* For any regex r, string s *)
  forall r m s, compileSubPattern r = Success m ->
    earlyErrors r = OK ->
    (* the matcher cannot fail an assertion. *)
    m (init_state s) (identity_cont) <> Failure.
\end{lstlisting}%
\end{minipage}
\Description{}
\caption{Mechanized theorems for termination and absence of failures.}%
\label{fig:coq_properties}
\end{figure}

\subsection{Proving Termination and Safety}%
\label{subsec:property_proofs}

Proving termination and safety (absence of failures) then amounts to proving the theorem of \cref{fig:invariant_preserved}.
This theorem can be proved by induction on regexes.
In this section, we describe some of the interesting cases.

\paragraph{Termination of the RepeatMatcher}

To compile quantifiers, the ECMAScript standard defines a function \inlinecode{RepeatMatcher}, which takes as argument a matcher function for the inner regex, and returns a matcher function for the quantified regex.
We show in \cref{fig:simpl-repeat-matcher} a simplified version of our mechanization of this \inlinecode{RepeatMatcher} function, for the special case of matching the quantifier $e\rrep{\texttt{min}}{}$, where \inlinecode{m} is the matcher function for $e$.
In essence, this function computes a new continuation \inlinecode{d}, which repeats the matcher \inlinecode{m} as many times as possible and at least \inlinecode{min} times.
The first \inlinecode{min} repetitions are allowed to match the empty string, but subsequent repetitions are required to make some progress in the string; otherwise \inlinecode{endIndex(y) = endIndex(x)} and the continuation returns without performing any additional recursive calls.

\begin{figure}
\begin{lstlisting}[language=Coq,breaklines=true,basicstyle=\figcodesize]
Definition RepeatMatcher (m: Matcher) (min: nat) (x: MatchState) (c: MatcherContinuation) (fuel: nat) :=
  match fuel with
  | 0 => out_of_fuel
  | S fuel' =>
    let d := fun (y: MatchState) =>
      if min = 0 and endIndex(y) = endIndex(x) then mismatch
      else
      let nextmin := if min = 0 then 0 else min - 1 in
      RepeatMatcher m nextmin y c fuel'
    in
    if min != 0 then m x d
    else
    let z := m x d in
    if z is not mismatch then z
    else c x
  end.
\end{lstlisting}%
\Description{}
\caption{Simplified pseudocode of \inlinecode{RepeatMatcher}~\cite[22.2.2.3.1]{ecma_262}.}%
\label{fig:simpl-repeat-matcher}
\end{figure}

\renewcommand{\min}[0]{\textup{min}}
\newcommand{\remchars}[1]{\textup{remainingChars}(#1)}
\newcommand{\fuel}[0]{\textup{fuel}}
\newcommand{\eiState}[1]{\textup{endIndex}(#1)}
\newcommand{\inputString}[1]{\textup{input}(#1)}
\newcommand{\stringLength}[1]{\textup{length}(#1)}

Recall (from \cref{subsec:termination}) that we used fuel to define this function in Coq, as the function is not structurally recursive.
Whenever we call this function, we call it with an extra argument, its fuel.
To be able to prove by induction on the regex the theorem of \cref{fig:invariant_preserved}, in the particular case where the regex is a quantifier, we have to prove as an intermediate theorem that we can compute an initial fuel value to give to the \inlinecode{RepeatMatcher} such that it does not run out of fuel.
We then state and prove the intermediate lemma described in \cref{fig:repeatmatcher_terminates}, using the bound
$\min{} + \remchars{x} + 1$
as initial fuel, where $\remchars{x}$ is the number of characters in the string which were not yet consumed in the match, i.e.
$\remchars{x} = \stringLength{\inputString{x}} - \eiState{x} \geq{} 0$.

Informally, this bound comes from the fact that each time the \inlinecode{RepeatMatcher} iterates, either the value of \inlinecode{min} has decreased, or the matcher has made some progress in the string (because the matcher invariant holds for \inlinecode{m}).
In the latter case, either \inlinecode{m} has not matched any character, and the \inlinecode{RepeatMatcher} immediately terminates, or \inlinecode{m} has matched some characters and the number of remaining characters decreases.

\begin{figure}
\begin{lstlisting}[language=Coq,basicstyle=\figcodesize]
Lemma repeat_matcher_terminates :
  forall m x c min,
    (* for any continuation c that terminates *)
    (forall y, c y <> OutOfFuel) ->
    (* for any matcher m with the invariant *)
    matcher_invariant m ->
    (* the matcher constructed by RepeatMatcher terminates *)
    RepeatMatcher m min x c (min+remainingChars(x)+1) <> OutOfFuel.
\end{lstlisting}%
\Description{}
\caption{Mechanizing termination of the RepeatMatcher.}%
\label{fig:repeatmatcher_terminates}
\end{figure}

\subsubsection{Absence of Failures}
When proving that the matcher invariant holds for any matcher function generated by the specification, we also need to prove that each assertion holds.
We highlight here two interesting cases.
\begin{itemize}
\item
  Surprisingly, the notion of progress in the matcher invariant is not only helpful for the termination of \inlinecode{RepeatMatcher}, but is also useful to show that assertions cannot fail.
  For instance, this allows to prove that the assertion at step \sstep{3.c.vi} of the compilation of capture groups~\cite[Section 22.2.2.7]{ecma_262} holds, since the assertion tests that progress has been made in the string.
\item
  When compiling a named backreference, the compilation function first finds all named capturing groups using that name, then asserts that this list contains exactly one element.
  \begin{minipage}{\linewidth}
  \begin{lstlisting}[language=Coq,basicstyle=\figcodesize]
(** >> AtomEscape :: k GroupName <<*)
| AtomEsc (AtomEscape.GroupEsc gn) =>
  (*>> 1. Let matchingGroupSpecifiers be GroupSpecifiersThatMatch(GroupName). <<*)
  let matchingGroupSpecifiers := groupSpecifiersThatMatch self ctx gn in
  (*>> 2. Assert: matchingGroupSpecifiers contains a single GroupSpecifier. <<*)
  assert! (List.length matchingGroupSpecifiers =? 1);
  \end{lstlisting}
  \end{minipage}

  This assertion holds because the early-errors phase rules out regexes that define the same named capture group twice and regexes with a backreference to an undefined capture group.
  As a result, we proved that compilation always succeeds when the early errors phase succeeds, as follows:

  \begin{minipage}{\linewidth}
  \begin{lstlisting}[language=Coq,basicstyle=\figcodesize]
Theorem early_errors_compile :
  (* For any regex without early errors *)
  forall r, earlyErrors r = OK ->
    (* compiling it to a matcher is a success *)
    exists m, compileSubPattern r = Success m.
  \end{lstlisting}
  \end{minipage}
\end{itemize}

\subsection{Generalizing Progress}%
\label{subsec:generalize_progress}

Up to this point, our definition of the matcher invariant assumed that regex matcher functions always progress forward in the string.
The proof we presented for the matcher invariant, from which both termination and absence of failures followed, crucially relied on this property.

However, JavaScript regexes include lookbehinds, which traverse the string backwards.
In the actual specification, the compilation functions take an additional parameter indicating the direction of the compiled matcher, i.e.\ whether it should go through the input string forward or backward.
This is notably used when compiling lookbehinds: when compiling $\lookbehind{r}$, $r$ is compiled with the direction set to \inlinecode{backward}, regardless of the direction used when compiling the lookbehind.
In our mechanization, we use a generalized notion of progress $\progress$ that it is parameterized by a direction.
In the context of the termination of \inlinecode{RepeatMatcher} (\cref{subsec:property_proofs}), the definition of $\remchars{x}$ is also parametrized on the direction.
Since the repeated matcher always goes in the same direction, our termination argument holds whether the matcher goes forward or backward.

\section{Formally Verifying an Optimization Using our Mechanization}%
\label{sec:strictly_nullable}

Mechanized semantics open the door to new research.
In this section, we illustrate this claim with a case study showing that our mechanization can be used to formally verify a regex optimization used in a widely deployed JavaScript implementation, Irregexp.

\def\SN{\mathit{SN}}

As noted in \cref{subsec:js_formalizations}, subtleties of JavaScript regex semantics invalidate many natural and intuitive regex transformations that engines could use to simplify regexes.
V8's Irregexp engine nonetheless performs regex rewriting during parsing:
it replaces $r\star{}$ with the empty regex $\epsilon$ whenever $r$ can only match the empty string.\footnote{\url{https://github.com/v8/v8/blob/dd8d0b44d5d5b4bc3912026bf78f25ed3cafda1a/src/regexp/regexp-parser.cc\#L3201}}
In this section, we build upon our Coq mechanization of the ECMAScript regex semantics to provide a Coq proof that this optimization is correct.
Informally, this transformation is valid because iterations of a star that do not consume any character in the string should be rejected by the star, meaning that the only acceptable behavior of the star is to perform zero iterations.

Our proof is conducted in four steps.
We first define a syntax-directed analysis of regexes, identifying regexes that can only match the empty string, which we call \textit{strictly nullable regexes}.
This analysis is depicted on \cref{fig:strictly_nullable_analysis}, where $\SN~r$ means that the regex $r$ is strictly nullable.
We include any anchors and lookaround. The definition is recursive for disjunction, concatenation, quantifiers and groups.
Note that strictly nullable regexes can still look at the surrounding string with lookarounds or anchors; lookarounds can even be used to capture non-empty strings, despite not consuming any character from the input.

\begin{figure}
\mbox{\infer[]{\SN~\epsilon}{}}%
\quad\mbox{\infer[]{\SN~\mstart{}}{}}
\quad\mbox{\infer[]{\SN~\mend{}}{}}
\quad\mbox{\infer[]{\SN~\wboundary{}}{}}
\quad\mbox{\infer[]{\SN~\wBoundary{}}{}}

\vspace{.5cm}
\mbox{\infer[]{\SN~\lookahead{r}}{}}
\quad\mbox{\infer[]{\SN~\neglookahead{r}}{}}
\quad\mbox{\infer[]{\SN~\lookbehind{r}}{}}
\quad\mbox{\infer[]{\SN~\neglookbehind{r}}{}}

\vspace{.5cm}
\mbox{\infer[]{\SN~r_1\disj{}r_2}{\SN~r_1\quad\SN~r_2}}
\quad\mbox{\infer[]{\SN~r_1 r_2}{\SN~r_1\quad\SN~r_2}}

\vspace{.5cm}
\mbox{\infer[]{\SN~r\rrep{n}{}}{\SN~r}}
\quad\mbox{\infer[]{\SN~r\rrep{n}{m}}{\SN~r}}
\quad\mbox{\infer[]{\SN~r\lrrep{n}{}}{\SN~r}}
\quad\mbox{\infer[]{\SN~r\lrrep{n}{m}}{\SN~r}}

\vspace{.5cm}
\mbox{\infer[]{\SN~\pgroup{r}}{\SN~r}}
\quad\mbox{\infer[]{\SN~\group{r}}{\SN~r}}
\quad\mbox{\infer[]{\SN~\group[name]{r}}{\SN~r}}

\caption{Our strictly nullable analysis}%
\Description{}
\label{fig:strictly_nullable_analysis}
\end{figure}

In our second step, we define a property of the matcher functions of such strictly nullable regexes.
This property encodes the fact that such matcher functions cannot make any progress in the string.
This can be seen as a stronger version of the matcher invariant mechanization of \cref{fig:matcher_invariant}, where the \inlinecode{endIndex} inequality has been replaced by an equality.
This is defined on \cref{fig:strictly_nullable_matcher}.

\begin{figure}
\begin{minipage}{.55\textwidth}%
\begin{lstlisting}[language=Coq,basicstyle=\figcodesize]
Definition strictly_nullable_matcher (m: Matcher) :=
(* For any valid state, continuation c *)
forall x c, Valid x ->
  (* either there is no match *)
  (m x c = mismatch) \/
  (* or m produced a valid state y which *)
  (exists y, valid y /\
    (* did not make progress *)
    endIndex x = endIndex y /\
    (* and was passed to c. *)
    c y = m x c).
\end{lstlisting}%
\caption{Strictly Nullable Matcher definition.}%
\label{fig:strictly_nullable_matcher}%
\end{minipage}%
\begin{minipage}{.45\textwidth}%
\begin{lstlisting}[language=Coq,basicstyle=\figcodesize]
Theorem strictly_nullable_analysis_correct:
  forall (r:Regex) (m:Matcher),
    strictly_nullable r = true ->
    compileSubPattern r = Success m ->
    strictly_nullable_matcher m.
\end{lstlisting}%
\Description{}
\caption{Correctness of the $\SN$ analysis.}%
\label{fig:strictly_nullable_correct}
\end{minipage}
\end{figure}

In a third step, we can prove that every strictly nullable regex, according to the analysis of the first step, is compiled to a strictly nullable matcher according to our new definition.
This theorem is shown in \cref{fig:strictly_nullable_correct}.
Its proof proceeds by induction on the regex.
It closely resembles the proof of the theorem of \cref{fig:invariant_preserved}, except that we use the fact that the regex is strictly nullable to show that the intermediate state \inlinecode{y} on which the continuation is called has made no progress in the string.

Finally, we can conclude that it is correct to replace $r\star{}$ with the empty regex when $r$ is strictly nullable.
We prove the theorem in \cref{fig:strictly_nullable_theorem}, showing that both $r\star{}$ and the empty regex are compiled to equivalent matchers, returning the same results when called with the same inputs.

\begin{figure}
\begin{lstlisting}[language=Coq,basicstyle=\figcodesize]
Theorem strictly_nullable_same_matcher:
  forall (r:Regex) (mstar: Matcher) (mempty: Matcher),
    strictly_nullable r = true ->
    compileSubPattern (Quantified r (Greedy Star)) = Success mstar ->
    compileSubPattern Empty = Success mempty ->
     forall x c, mstar x c = mempty x c.
\end{lstlisting}%
\Description{}
\caption{Correctness of the strictly nullable optimization}%
\label{fig:strictly_nullable_theorem}
\end{figure}

Even simple, seemingly intuitive transformations can be incorrect in JavaScript regexes (see \cref{subsec:js_formalizations}), and our Coq mechanization enables the formal verification of regex engine optimizations.
One direction of possible future work could be to make the strictly nullable analysis even more precise.
Currently, backreferences are never considered strictly nullable, and this is the case in both our analysis of \cref{fig:strictly_nullable_analysis} and in the V8 optimization.\footnote{\url{https://github.com/v8/v8/blob/74530c3ff422d6e10ece2d9b68f0caec74c94a63/src/regexp/regexp-ast.h\#L721}}
But when a backreference refers to a capture group whose body is itself strictly nullable, then the backreference itself should be strictly nullable.
We believe that our analysis and proof could both be extended to justify even more aggressive optimizations.

\section{Evaluation}
\label{sec:evaluation}

We claim that our mechanization of ECMAScript regexes is proven-safe, executable, faithful and future-proof.
\Cref{sec:proofs} describes our mechanization: we have formally proved that the specification always terminates and cannot fail.
We substantiate the claims of executability, faithfulness, and robustness to future changes in sections~\ref{subsec:extraction}, \ref{subsec:faithful}, and~\ref{subsec:future-proof}.
Finally, we discuss the formalization effort in section~\ref{subsec:effort}.

\subsection{An Executable Mechanization}%
\label{subsec:extraction}

Using Coq's extraction mechanism~\cite{coq_extraction}, we can generate OCaml code from our mechanization.
This OCaml code can then be compiled, linked with adequate libraries, and executed to produce an independent engine for matching JavaScript regexes.
Separately, to better integrate with the JavaScript ecosystem, we use Melange~\cite{melange} to translate the generated OCaml code to JavaScript.

As a result, from our Coq mechanization we can automatically generate two executable regex engines, one in OCaml, and one in JavaScript.
Below, we highlight some interesting aspects of the extraction, compilation, and linking process.

\paragraph{Parsing}
ECMA-262 specifies the syntax of regular expressions, but we did not port this part to Coq.  Instead, we use an existing JavaScript regex parser library, \inlinecode{regexpp}~\cite{regexpp} to parse regex strings into ASTs, and a small amount of code of JavaScript code to translate these ASTs into the representation used by our mechanization.
This choice is motivated by a divergence within the ECMAScript standard, which defines two different grammars (one in section 22.2.1, and one in annex B.1.2).
The former is the standard grammar, and the latter is the ``legacy'' grammar, intended for web browsers but used by most implementations.
These grammars do not agree on all inputs: for instance, the standard grammar rejects \inlinecode{]}, whereas the legacy grammar allows it and treats it  as a valid regex matching the character `\ch{]}'.
Given this, we left parsing out of our mechanization.

\paragraph{Unicode support}
JavaScript regexes support Unicode through the \inlinecode{u} flag.
In JavaScript, strings are represented as sequences of UTF-16 code units.
The presence or absence of the \inlinecode{u} flag affects their tokenization during matching~\cite[Section 22.2.2]{ecma_262}.
Without the \inlinecode{u} flag, the input string is processed as-is, treating each code unit as one character.
With the \inlinecode{u} flag, the input string is tokenized into a sequence of Unicode characters before processing.
As an example, `\includegraphics[width=.4cm]{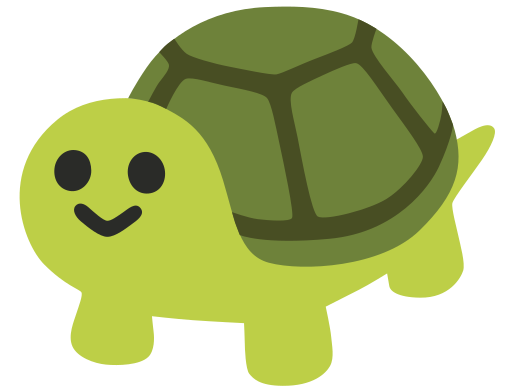}' is a single Unicode character, \inlinecode{U+1F422}, but is encoded as a sequence of two code units (\inlinecode{0xD83D} \inlinecode{0xDC22}) in UTF-16 (and hence in JavaScript strings).
Matching this string against the regex $\dotc{}$ (which matches a single token) produces different results depending on the Unicode flag.
When the flag is active, the regex matches the whole input, since the two code units get tokenized into just one Unicode character.
Otherwise, the regex matches only the first code unit, \inlinecode{0xD83D}: the engine considers that the string contains two distinct tokens.

In our mechanization, we reflect these tokenization choices by parameterizing the whole development on a type representing characters, along with functions to manipulate them (e.g., case foldings for case-insensitive matching).  This means, in particular, that our specification does not include a specification of all Unicode: we delegate Unicode operations to external code that links against our specification.

\begin{figure}
\centering
\includegraphics[width=\textwidth]{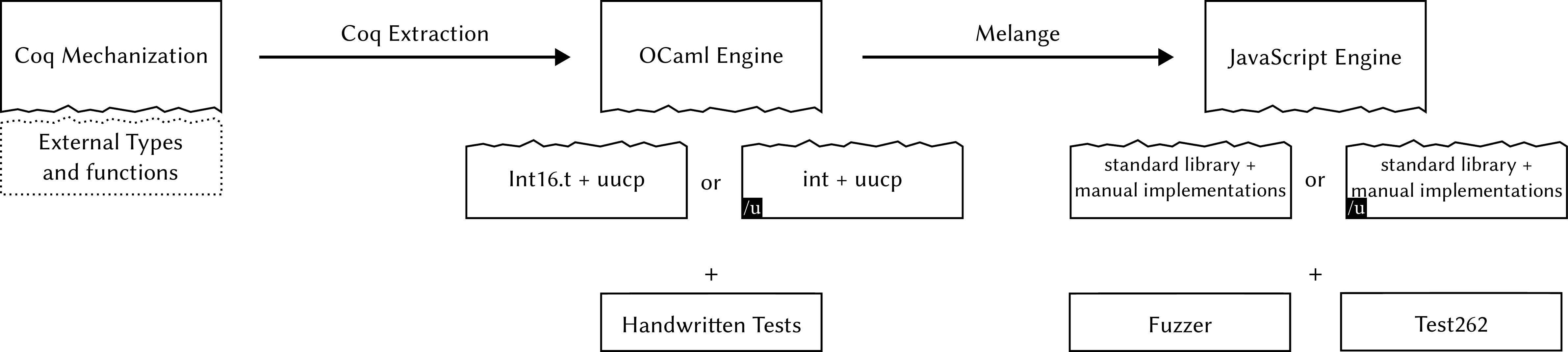}
\Description{A diagram with three compilation steps: A Coq mechanization parameterized by external types and functions, extracted to an OCaml Engine, compiled with Melange to a JavaScript Engine.  The OCaml and JavaScript engines are linked with implementations based on uucp or the JS standard library, and tested with handwritten tests or a fuzzer and Test262.}
\caption{Architecture of our development.}
\label{fig:archi}
\end{figure}

We then provide two ways to instantiate the specifications, one corresponding to regexes with the \inlinecode{u} flag, and one without.
The entire architecture of our development is explained in \cref{fig:archi}, where both OCaml end JavaScript engines are completed with implementations of these types and functions: one for non-Unicode mode, and one for Unicode mode.

In the OCaml engine without Unicode mode, characters are encoded with the \inlinecode{Int16.t} type from the \inlinecode{integers} library.
In the OCaml engine with Unicode mode, characters use the \inlinecode{int} type (large enough to represent all Unicode code points), and we use the OCaml library \inlinecode{uucp} to implement character manipulation functions like case foldings.

In the JavaScript engine, we use JavaScript strings to represent characters in all cases.  We use standard library functions to implement some of the axiomatized character-manipulation operations (such as changing to uppercase), and manual implementations of the remaining ones (such as Unicode case foldings).

In other words: for case foldings or Unicode properties, the ECMAScript specification implicitly depends on an independent standard: Unicode.
Our Coq mechanization abstracts the corresponding definitions away.
To demonstrate the practicality of this approach, we instantiated most of these functions in our OCaml and JavaScript engines using existing Unicode libraries and manual implementations: we have support for all character manipulation functions, and we added support for one of the \texttt{\escaped{p}} property classes as a sanity check.

\subsection{A Faithful Mechanization}%
\label{subsec:faithful}

Our mechanization is trustworthy not just because we put great care into following the specification to the letter, but because we made every effort to validate it empirically.
To do so, we used three layers of testing: we started with 95 custom tests, created by hand as we developed the mechanization, which we executed directly on the extracted OCaml code.  Then, we compiled the extracted code to JavaScript, and made sure that the result passed the regex test suite used by most web browser vendors, Test262.  Finally, to gain additional confidence, we subjected the same JavaScript code to hours of differential fuzzing against Irregexp, the regex engine found in V8.

\paragraph{Differential fuzzing}
We implemented a differential fuzzer to cross-validate the correctness of our implementation with the Irregexp implementation from V8~\cite{v8}, using randomly generated regexes and strings.
The regexes and strings are generated on a small restricted alphabet of a few characters to improve the chances of finding matches.
Our tests use randomly selected flags and start at random positions in the string.
When comparing the results of our extracted engine and of Irregexp, we compare the value of the whole match and each capture group.

In this process, we directly generate regex ASTs, meaning that our fuzzer does not test the parser but rather the three phases of \cref{subsec:matching} that we mechanized in Coq.
Over thousands of such random tests, our extracted JavaScript regex engine has never disagreed with Irregexp.

\paragraph{Executing the Test262 suite}
Test262~\cite{test_262} is the official ECMAScript test suite, covering all of JavaScript, and in particular regexes.
To ensure that we faithfully mechanized the ECMAScript regex specification, we made sure that our extracted JavaScript engine passed the regex tests included in Test262.

Doing so required careful plumbing, because Test262 regex tests are not plain input-output pairs: they are complete JavaScript programs that call regex API functions and assert properties of the results.  As such, these tests must be run with a full JavaScript engine: they cannot be run with a standalone regex engine.
To validate our mechanization, we ran them with V8, but in a modified environment set up to route all calls to \inlinejs{RegExp.prototype.exec} through our own code.
In this set up, any time a test calls \inlinejs{exec}, execution transfers to our code, and we can use \texttt{regexpp} to parse the given regex, forward the parsed regex and the input to our own JavaScript regex engine, and finally return the results to the caller, at which point normal V8 execution resumes using the result computed by our engine.
This way, we can check that our mechanization agrees with the tests of Test262.

\paragraph{Relevant tests}
Test262 contains 498 tests of core regex-matching functionality. They contain assertions like \inlinejs{assert.compareArray("abcdef".match(/(?<=ab(?=c)\\wd)\\w\\w/), ["ef"], "#1");}\footnote{\url{https://github.com/tc39/test262/blob/c95cc6873d9933b8674ec8a840f43be0fb836ec1/test/built-ins/RegExp/lookBehind/nested-lookaround.js\#L24}}, which checks the result of a call to \inlinejs{match} with nested lookarounds. These tests are directly relevant to our Coq mechanization.
We pass 495 out of 498; the remaining 3 tests time out.
The three tests that time out are character classes tests for the regexes \escaped{D}, \escaped{W} and \escaped{S} which are checked exhaustively by testing all Unicode characters one by one. The algorithm described by the specification iterates over the entire character class for each input character, and since \escaped{D}, \escaped{W}, and \escaped{S} include almost all Unicode characters, each test ends up executing billions of comparisons.

\paragraph{Other tests}
Test262 also includes other tests that involve JavaScript regexes yet are not directly relevant to the parts of the semantics we mechanized in Coq. We detail them below:

\begin{itemize}
\item There are 445 prototype tests that check whether JavaScript objects related to regex matching have the right properties.
For instance,
{\begin{center}\inlinejs{assert.sameValue(RegExp.hasOwnProperty('prototype'), true, ...)}\footnote{\url{https://github.com/tc39/test262/blob/c95cc6873d9933b8674ec8a840f43be0fb836ec1/test/built-ins/RegExp/prototype/S15.10.5.1_A1.js}}\end{center}}
checks the presence of the \inlinejs{prototype} JavaScript property on the Regexp object.
We can still run our extracted JavaScript regex engine on these tests, and we pass 436 out of 445 tests.
Of the 9 tests that we fail, 5 are tests that fail because of V8.\footnote{\url{https://issues.chromium.org/issues/42203113} and \url{https://issues.chromium.org/issues/42200389}}
The remaining 4 fail because of the JavaScript wrapper function that we use to intercept V8 and switch to our engine.
For instance, one of the tests removes some string and array methods from the standard library and checks that regex matching still works\footnote{\url{https://github.com/tc39/test262/blob/c95cc6873d9933b8674ec8a840f43be0fb836ec1/test/built-ins/RegExp/prototype/Symbol.replace/poisoned-stdlib.js\#L10}}, but our JavaScript wrapper uses some of these methods.

\item There are 560 Unicode property-class tests that are mostly auto-generated~\cite{prototype_tests} and test each individual character property class exhaustively. These are orthogonal to our mechanization: property classes are defined in a separate standard, the Unicode Character Database (UCD), and are handled by external Unicode libraries. Although we added one such property class as a sanity check, the corresponding tests would time out just like the ones for \escaped{D}, \escaped{W}, and \escaped{S} above, so we excluded them.

\item Finally, Test262 also contains 192 tests about experimental features currently under review for future inclusion in the specification. For example,
{\begin{center}\inlinejs{assert.compareArray("bab".match(/(?<x>a)|(?<x>b)/), ["b", undefined, "b"]);}\footnote{\url{https://github.com/tc39/test262/blob/c95cc6873d9933b8674ec8a840f43be0fb836ec1/test/built-ins/RegExp/named-groups/duplicate-names-match.js\#L11}}\end{center}}
tests support for duplicate named groups, which are currently forbidden by the specification.
As our mechanization is up to date with the current published specification, it does not pass these tests.
\end{itemize}

\subsection{A Future-Proof Mechanization}%
\label{subsec:future-proof}

The ECMAScript specification receives frequent updates, and so does the regex chapter.
Our mechanization can easily be modified as the prose ECMAScript standard evolves.
We claim that the changes required in our mechanization should be of the same order of magnitude as the changes to the specification itself.

As a case study, we delayed the implementation of word boundary assertions, or \emph{anchors}, ($\mstart{}$, $\mend{}$, $\wboundary{}$ and $\wBoundary{}$), to the very end of our mechanization effort, so we could simulate the addition of a simple feature to the specification.
In this section, we document the changes that were needed to accommodate the introduction of anchors into the specification.

The first step is to extend the \inlinecode{Regex} type to add the four new constructors (one for each boundary assertion).
Since these constructors are leaves of the AST, there is no need to extend the zipper with new context constructors.

Because we introduced new AST constructors, we are also required to update some methods of section 2.2.1 such as \inlinecode{CountLeftCapturingParensBefore} as well as \inlinecode{CompileSubPattern}, the main compilation function.
Next, we need to implement the new cases in \inlinecode{CompileSubPattern} to compile boundary assertions.
This process is made straightforward by our notations: most of the pseudocode can be transcribed almost literally, as shown in \cref{fig:case-study-b}.
The most difficult line to translate is the one for case \inlinespec{1.e}.
The condition requires indexing into the input string, which is an operation whose result is wrapped in the error monad as explained in \cref{sec:error-monad}.
As such, its result must be unwrapped using \inlinecode{let!} before passing it to \inlinecode{CharSet.contains}.
We defined new boolean monadic notations \inlinecode{if!}, \inlinecode{\&\&!} and \inlinecode{||!}\hspace{-.15cm} to encode such potentially failing conditions.

\begin{figure}
\begin{lstlisting}[language=Coq,breaklines=true,basicstyle=\figcodesize]
(** >> Assertion :: ^ <<*)
| InputStart =>
    (*>> 1. Return a new Matcher with parameters (x, c) that captures rer and performs the following steps when called: <<*)
    (fun (x: MatchState) (c: MatcherContinuation) =>
      (*>> a. Assert: x is a MatchState. <<*)
      (*>> b. Assert: c is a MatcherContinuation. <<*)
      (*>> c. Let Input be x's input. <<*)
      let input := MatchState.input x in
      (*>> d. Let e be x's endIndex. <<*)
      let e := MatchState.endIndex x in
      (*>> e. If e = 0, or if rer.[[Multiline]] is true and the character Input[e - 1] is matched by LineTerminator, then <<*)
      if! (e =? 0)%Z ||! (
            (RegExp.multiline rer is true) &&!
            (let! c =<< input[(e-1)%Z] in CharSet.contains CharSet.line_terminators c))
      then
        (*>> i. Return c(x). <<*)
        c x
      else
      (*>> f. Return failure. <<*)
      failure): Matcher
\end{lstlisting}%
\Description{}
\caption{Extension of \inlinecode{CompileSubPattern} for the $\wboundary$ construct.}%
\label{fig:case-study-b}
\end{figure}

The changes above are enough to fully support boundary assertions.
We also adapted the proofs described in \cref{sec:proofs} to these new constructs.
As boundary assertions do not affect termination and do not have complex assertions, this only required a few additional lines of proof.

A similar process, albeit at a larger scale, should make it possible to support features added in the 2024 edition of the ECMAScript specification, or currently under discussion,
such as the \texttt{v} flag\footnote{\url{https://github.com/tc39/proposal-regexp-v-flag}},
duplicated named groups\footnote{\url{https://github.com/tc39/proposal-duplicate-named-capturing-groups}},
pattern modifiers\footnote{\url{https://github.com/tc39/proposal-regexp-modifiers}}, etc.

\subsection{Formalization Effort}
\label{subsec:effort}

The ECMAScript specification contains around 33 pages dedicated to regex matching, which we mechanized in Coq.
The table below summarizes our formalization effort.
Additional handwritten tests were written alongside the mechanization.

\begin{center}
\begin{tabular}{l l r}
        Language & Category & Lines of Code \\
        \hline\hline
        Coq & Semantic mechanization & 1810 \\
        Coq & Infrastructure and type definitions & 788 \\
        Coq & Termination and safety proofs & 1392 \\
        Coq & Strictly nullable optimization proofs & 477 \\
        OCaml & Extracted engine infrastructure & 688 \\
        OCaml & OCaml parameter implementation & 244 \\
        JavaScript \& OCaml & JavaScript parameter implementation & 255 \\
        OCaml & Handwritten tests & 973 \\
        JavaScript \& OCaml & Differential fuzzer & 374 \\
        JavaScript \& OCaml & Test262 wrapper & 137 \\
\end{tabular}
\end{center}

\section{Limitations and Future Work: Looking Ahead}
\label{sec:future_work}

We have formalized the ECMAScript regex matching process in its entirety, but not the external data that it depends on (Unicode property tables and character manipulation functions) nor the APIs that expose it.

In our mechanization, we have included as a comment, for each line of Coq code, the corresponding pseudocode line from the ECMAScript specification.
To ensure completeness, we could build a tool to extract these comments and check that each and every line of the ECMAScript standard has been translated.

\paragraph{Formal verification of regex engines}
As future work, we believe that our mechanized semantics would constitute the ideal specification for the correctness theorem of a formally verified JavaScript regex engine.
Some JavaScript engines employ techniques fundamentally different from backtracking to avoid the particular cases where backtracking has exponential time complexity.
For instance, both the Experimental engine in V8~\cite{non_backtrack_v8} and recent work on linear matching of JavaScript regexes~\cite{linear_matching_js} use a technique known as NFA simulation, which explores several paths in parallel.
This approach is so different from backtracking that subtle semantic bugs have been uncovered~\cite{experimental_bug}.
Proving the correctness of this approach using our mechanization as a specification would provide high assurance that engines with linear-time complexity can be used to match JavaScript regexes.

Backtracking implementations sometimes use more advanced optimizations than the one we formally verified in \cref{sec:strictly_nullable}.
For instance, Irregexp generates native code when the same regex is being executed several time~\cite{v8_tier_up}, and recent work has explored the use of memoization to speed up backtracking engines even with modern extended regex features such as lookarounds or backreferences~\cite{selective_memo}.
One could also use our mechanization to formally verify these techniques in Coq, for instance proving the correctness of a native code compiler for JavaScript regexes, or proving the correctness of an efficient memoized backtracking implementation.

Another kind of engine uses regex derivatives.
Derivatives~\cite{brzozowski64} present an elegant way to define the semantics of standard regular expressions.
A recent addition to the .NET framework is a linear-time derivatives-based engine for a subset of .NET regexes (without backreferences or lookarounds)~\cite{dotnet_pldi}.
To our knowledge, no derivatives-based engine exists for JavaScript regexes.
One crucial step in defining a derivatives-based engine is to use regex rewriting rules to group together semantically equivalent regexes.
In essence, for standard regular expressions, this ensures that the algorithm explores a \textit{finite} equivalent automaton.
A list of such rewriting rules for standard regular expressions (extended with regex intersection, complement and the empty-set regex) is available in~\cite[Section 4.1]{derivatives_reexamined} and combining these rules has been proven to make the set of derivatives finite up to equivalence.
However, for JavaScript regexes, we have already used our executable engine to show that several of these rules are incorrect in JavaScript, as shown on \cref{fig:transformations}.
For the first rule, deleting branches may change the index of capture groups. For the second one, disjunction is notoriously non-commutative in JavaScript because the left branch has higher priority than the right one.
This raises a new open question: is it possible to find a combination of rewriting rules leading to the same finiteness result for JavaScript regexes?
If there is, we believe we could use our mechanization to formally verify each transformation, just like we did in \cref{sec:strictly_nullable}.

\begin{figure}
\begin{tabular}{r c l@{\hskip 1in}l}
\midrule\multicolumn{4}{c}{Derivative equivalence rules~\cite{derivatives_reexamined}}\\ \midrule
\multicolumn{2}{c}{Rewriting Rule}& & JavaScript Counterexample\\
$r\disj{}r$ &$\equiv$& $r$ & $\pgroup{\group{a}\disj{}\group{a}}\nbackref{2}()\$$ on string \str{aa}.\\
$r_1\disj{}r_2$ &$\equiv$& $r_2\disj{}r_1$ & $\ch{a}\disj{}\ch{a}\ch{b}$ on string \str{ab}.
\end{tabular}

\vspace{.5cm}
\begin{tabular}{r c l@{\hskip 1in}l}
\midrule\multicolumn{4}{c}{Rules from ~\cite{expose,regex_repair} (see \cref{subsec:js_formalizations})}\\ \midrule
\multicolumn{2}{c}{Rewriting Rule}& & JavaScript Counterexample\\
$r\question{}$ &$\equiv$& $r\disj{}\epsilon$ & $()\question{}$ on the empty string \str{}.\\
$r\lquestion{}$ &$\equiv$& $\epsilon\disj{}r$ & $\lookahead{\group{a}}\lquestion{}\ch{a}\ch{b}\nbackref{1}\ch{c}$ on string \str{abac}.\\
\end{tabular}
\Description{}
\caption{Counterexamples to standard regular expression rewriting rules we found during our mechanization.}
\label{fig:transformations}
\end{figure}

\paragraph{Formal verification of other JavaScript regex semantics}
Previous work reasoning about JavaScript regexes~\cite{expose,regex_repair} have preferred a more denotational style of semantics, rather than mimicking the operational backtracking algorithm of the ECMAScript specification.
We believe that these models could benefit from our mechanization by being proved equivalent to it in Coq.
Similarly, the transducers semantics of~\citet{psst} have been tested against JavaScript, and with our mechanization we could prove that the two semantics are in fact equivalent.

\paragraph{Completing an existing JavaScript mechanization}
Our mechanization could be used to complete the JSCert~\cite{jscert} JavaScript mechanization in Coq.
To do so, we would also need to mechanize some regex API functions that call our \inlinejs{RegExp.prototype.exec} function.
As an experiment, we already mechanized the \inlinejs{search}, \inlinejs{test}, \inlinejs{match} and \inlinejs{matchAll} functions, and included these functions in our differential fuzzer.  Our specification style seemed to work just as well for these functions as for the regex matching specification.  We did not attempt to mechanize \inlinejs{replace} and \inlinejs{split}, because they call out to more generic functions about strings defined in other chapters of the ECMAScript specification.

\section{Related Work}%
\label{sec:related}

\paragraph{JavaScript mechanizations}
Several works have tackled the mechanization of the ECMAScript standard.
However, none of them covered the regex chapter.
The common explanation is that the style used to define the regex semantics is quite different from the style used in other chapters~\cite{tc39notes}.

For instance, the JSCert project~\cite{jscert,jsexplain} manually translated most of the ECMAScript 5 specification into Coq definitions.
Just like in our case, JSCert can then extract these Coq definitions to OCaml to obtain a reference interpreter, JSRef.
JSCert models the complex control-flow of the ECMAScript semantics using pretty-big-step semantics~\cite{prettybigstep}, and also uses monadic operators to encode possibly failing operations.
Our monadic definitions are more specialized to the kind of failures that occur in the regex chapter.
An interesting direction for further work would consist in combining the JSCert development with our regex mechanization.

Instead of manually translating ECMAScript pseudocode, the ESMeta project uses JISET, an automatic translator, to transpile ECMAScript pseudocode into a custom intermediate representation~\cite{jiset}.
This allows the mechanization to be updated almost automatically when the ECMAScript standard evolves.
This intermediate representation can then be used to define differential testers~\cite{jest} and static analysers~\cite{jstar,automatic_sa} for JavaScript.
Despite JISET being able to translate more than 90\% of the ECMAScript standard automatically, it also does not support the regex chapter (again, this is due to the peculiarities of the regex section).  Unfortunately, it also does not yet support generation of definitions suitable for use in an interactive theorem prover such as Coq.
To be able to rebuild our work on top of ESMeta, we would need to extend the JISET intermediate representation and translator to work with the regex specification style, and to add a translator from it to Coq; this could constitute interesting future work, and we hope that this paper would provide a good ground truth to evaluate against.

\paragraph{Regex formalizations}
Other than related work formalizing JavaScript regexes~\cite{expose,regex_repair} (see \cref{subsec:js_formalizations}),
there have been numerous formalizations of other regex languages, with various combinations of features.
We present a selection of different semantic choices here, although the list is not exhaustive.

A formal model of a subset of .NET regexes has been presented in~\cite{dotnet_pldi}, where the semantics is defined using derivatives extended with locations to support features like anchors and lookarounds.
A similar approach has recently been mechanized in Lean~\cite{lean_lookarounds} for regexes with lookarounds and POSIX longest match semantics.
Using derivatives to define the semantics of lookahead had been previously investigated in~\cite{derivatives_lookahead}.
In contrast with JavaScript regexes, these models do not include capture groups and backreferences.

There have also been different definitions of regex semantics using transducers, either \textit{prioritized finite transducers} (PFTs)~\cite{semantics_wild,analyzing_catastrophic} and later \textit{prioritized streaming string transducers} (PSSTs)~\cite{psst}.
Using PSSTs has allowed not only to define semantics for regexes with capture groups, but also to define semantics for API functions like \inlinejs{replace} or \inlinejs{replaceAll}, albeit without support for lookarounds and backreferences.
In contrast, our mechanization does not cover API functions (see \cref{sec:future_work}), but covers a wider range of regex constructs.

The semantics proposed in~\cite{lookahead_backref} supports a combination of capture groups, lookaheads, and backreferences to prove theorems about the expressive power of combining these features.
Interestingly, this semantics diverges from JavaScript on the interaction of lookaheads and backreferences.
In JavaScript, lookaheads are \textit{atomic} (see~\cite[22.2.2.4, Note 3]{ecma_262}), meaning that if there are several ways to match a lookahead, only the first one will be tried and a backreference cannot cause the engine to backtrack inside a lookahead.

\paragraph{Regex semantics}
Differences between regex languages have been studied in various contexts.
A number of syntactic and semantic differences have been documented in~\cite{lingua_franca} for JavaScript, Java, PHP, Python, Ruby, Go, Perl and Rust regexes, with JavaScript being different from all others in several cases.
That study already presented the JavaScript peculiarities of \cref{subsubsec:comparison}, but not the first, third and fourth counterexamples of \cref{fig:transformations}.
Recently, work that presented algorithms for matching a large subset of JavaScript regexes in linear time~\cite{linear_matching_js} (albeit without mechanized proofs or a formal model) further showed that subtleties of quantifier semantics had caused bugs in V8.
The semantics of backreferences are another source of differences across languages.
Backreferences can either match the empty string or nothing when the corresponding group is undefined, and some languages can allow multiple definitions of named groups with the same name.
Recent work~\cite{backref_reexamined} showed the impact of these semantic choices on the expressivity of the language.
JavaScript backreferences correspond to their \textit{No-label repetitions, $\epsilon$-semantics} category.

\section{Conclusion}%
\label{sec:conclusion}

We have presented an executable, proven-safe, faithful and future-proof Coq mechanization of JavaScript regex semantics.
We have shown that with the right combination of encoding techniques, a manual shallow embedding of the ECMAScript backtracking algorithm is possible in Coq and can scale to the full specification.
We have additionally proved that the specification always terminate and that failures never occur, two properties that could not have been formally verified without a mechanization.
Our work can be used to conduct more proofs about JavaScript regex semantics, for instance proving the correctness of optimizations used in widely deployed implementations.
Finally, our development can be extracted to OCaml or JavaScript, providing two executable regex engines.
We validated the correctness of our mechanization with the Test262 official conformance test suite, and cross-validated it against the V8 Irregexp engine.
To our knowledge, this is the first time the ECMAScript specification of regexes has been mechanized in a proof assistant.

Our work lays the foundations for future formal work about real-world regex languages.
In particular, we hope that it will make it possible to prove the correctness of efficient matching algorithms, as well as rewriting-based optimizations for regexes.
We also hope that it will provide a foundation for researchers to restate the semantics in a way that better suits their field, without sacrificing the assurance that these new formulations correctly reflect the ECMAScript regex language.

\section*{Acknowledgments}

The authors would like to thank Steve Zdancewic, as well as the anonymous reviewers, for their suggestions and feedback.

This project was supported by the Open Research Data Program of the ETH Board.

\section*{Artifact Availability}

The mechanization and proofs described in this paper are free software.  They can be found online at \url{https://github.com/epfl-systemf/Warblre}.

\nocite{artifact_reviewed}
A peer-reviewed artifact~\cite{artifact} is also available.  It consists of a virtual machine with our Coq mechanization, our proof scripts, and our auxiliary code (fuzzer, tests), as well as scripts to recreate the virtual machine from scratch.

\bibliographystyle{ACM-Reference-Format}
\bibliography{main}
\end{document}